\documentclass[useAMS,usenatbib]{mn2e}
\usepackage{url}
\usepackage{supertabular}
\usepackage{subfig}
\usepackage{graphicx}
\usepackage[countmax]{subfloat}
\usepackage{multirow}
\usepackage{morefloats}
\usepackage{booktabs}
\usepackage{amsmath}
\usepackage{amssymb}

\begin{document}

\title{A Prescription for Galaxy Biasing Evolution as a Nuisance Parameter}


\author[L. Clerkin et al.]
  {L. ~Clerkin,$^1$ D. ~Kirk,$^1$ O. ~Lahav,$^1$ F. B. ~Abdalla,$^1$ E. Gazta\~{n}aga $^2$\\
  $^1$University College London, Gower Street, London WC1E 6BT, UKS\\
  $^2$Institut de Ci`encies de l’Espai (ICE, IEEC/CSIC), E-08193
Bellaterra (Barcelona), Spain}

 \maketitle

\begin{abstract}

There is currently no consistent approach to modelling galaxy bias evolution in cosmological inference. This lack of a common standard makes the rigorous comparison or combination of probes difficult. We show that the choice of biasing model has a significant impact on cosmological parameter constraints for a survey such as the Dark Energy Survey (DES), considering the 2-point correlations of galaxies in five tomographic redshift bins. We find that modelling galaxy bias with a free biasing parameter per redshift bin gives a Figure of Merit (FoM) for Dark Energy equation of state parameters $w_0, w_a$ smaller by a factor of 10 than if a constant bias is assumed. An incorrect bias model will also cause a shift in measured values of cosmological parameters. Motivated by these points and focusing on the redshift evolution of linear bias, we propose the use of a generalised galaxy bias which encompasses a range of bias models from theory, observations and simulations, $b(z) = c + (b_0 - c)/D(z)^\alpha$, where parameters $c, b_0$ and $\alpha$ depend on galaxy properties such as halo mass. For a DES-like galaxy survey we find that this model gives an unbiased estimate of $w_0, w_a$ with the same number or fewer nuisance parameters and a higher FoM than a simple $b(z)$ model allowed to vary in z-bins. We show how the parameters of this model are correlated with cosmological parameters. We fit a range of bias models to two recent datasets, and conclude that this generalised parameterisation is a sensible benchmark expression of galaxy bias on large scales.



\end{abstract}

\section{Introduction}
\label{intro}

Galaxy biasing describes how the distribution of galaxies traces  the underlying matter distribution. The clustering of large scale structure is a potentially powerful probe of cosmology but it suffers from the problem that many of our techniques, such as redshift surveys, measure the light from galaxies rather than the underlying matter distribution; so our ability to do cosmology with this data is only as good as our understanding of the galaxy bias. 

While galaxy biasing is an important focus of direct study (e.g. \cite{Benson2000}, \cite{Seljak2005}, \cite{Gaztanaga2005a}, \cite{Cresswell2009a}, \cite{McBride2011}, \cite{Zehavi2011}) it often appears as a nuisance parameter required for the inference of other cosmological parameters. A nuisance parameter is defined as any parameter in which one is not immediately interested, and such parameters are marginalised over in Bayesian parameter estimation. Their impact on the parameters inferred can be significant since correlations with nuisance parameters will inflate the errors in the parameters of interest, so it is important to treat them rigorously. Galaxy bias is a nuisance parameter required to constrain, for example, the dark energy equation of state (e.g. \cite{Tegmark2004}), which is the focus of this analysis.


The concept of galaxy bias came about when it was noticed that different populations of galaxies demonstrate different clustering strengths (e.g. \cite{Davis1976}, \cite{Dressler1980}, leading to the conclusion that they cannot all have the same relationship linking their distribution with that of the matter. A physical mechanism for galaxy bias, that galaxies would tend to form in peaks in the matter density distribution thus being more clustered than the underlying matter distribution, was suggested by \cite{Kaiser1984} and developed by \cite{Bardeen1986}. Early clustering measurements also indicate that galaxy bias can not be linear (e.g. see \cite{Gaztanaga1992}, \cite{Fry1993}. Indeed, a linear bias relation could not be preserved through the non-linear growth of structure, and analytic models (e.g. \cite{Mo1996} and N-body simulations (e.g. \cite{Guo2009} provide a description of this non-linearity. We also know that the true bias relation is likely to be stochastic \cite{Dekel1999} since it is not possible to specify the galaxy distribution without also specifying numerous `hidden variables' such as their luminosity, temperature, physical size \textit{etc}. which would cause some physical scatter in the relation between the galaxy and matter density fields. Additionally, stochasticity is introduced into the measurement because of the discrete samples of the density field selected. Galaxy biasing evolves with redshift \cite{Nusser1994}, \cite{Fry1996}, \cite{Tegmark1998}, being naturally larger at high redshift as the first galaxies to form would have done so in the densest regions. It is also scale dependent at small physical scales where the non-linear effects of galaxy formation are important, although this is weak on large scales (\cite{Mann1998}), being scale invariant to within a percent for physical scales $r>20$ Mpc $h^{-1}$ (\cite{Crocce2013a}. In the analysis that follows we restrict ourselves to such large scales and address purely the redshift dependence of galaxy biasing. The constants $\alpha, b_0$ and $c$ of the proposed biasing evolution model are however expected to be sensitive to characteristics such as luminosity, colour and type, and while we defer a full treatment of this to future work, in section \ref{mass_dependence} we demonstrate the dependence of these parameters on mass.



Even within the relatively simple regime that we focus on of linear scales and redshift evolution of the galaxy biasing, there is currently no consensus on which model of biasing should be used. There is a gap in the literature between detailed modelling of biasing over the past 30 years and its implementation in current cosmological inference from data. A common solution to the lack of certainty in the biasing relationship is to dispense with physical insight and use a linear biasing parameter per redshift bin, but there are several issues with this. An approach whose features are closely linked to the specifics of individual survey design and data analysis may be useful for a given measurement, but makes it almost impossible to consistently compare different results. For example, if one analysis has five redshift bins and another has seven, or the same number of bins over a different redshift range, how can the results be meaningfully compared or combined? And while a small number of biasing parameters for a photometric redshift survey seems reasonable, a spectroscopic survey with 40 redshift bins would require an excessive number of nuisance parameters. Finally, it is preferable to use a physical model both because it enables one to probe the physical mechanism of biasing and it ensures that the nuisance parameters used are well motivated. For some nuisance parameters, such as intrinsic alignment, we do not have a physical model so a free parameter or function is appropriate. But where we do have information about the physics involved, as is the case for galaxy biasing, using this to select a small number of physically motivated parameters makes sense. 

The aim of this paper is to present and justify a generalised parameterisation of galaxy bias, physically motivated and restricted to redshift dependence and linear theory, which is a sensible candidate for a much needed benchmark expression of galaxy biasing. In section \ref{bias:models} we introduce the bias models considered in our analysis. In section \ref{impact_on_forecasts} we address the question of whether we should care which bias model is used, or that there is no consensus: firstly by quantifying the effect that the choice of bias model has on parameter forecasts using Fisher matrices, and secondly by looking at the shift in estimated parameter values that is introduced by assuming an incorrect bias model. We also check for degeneracies of the proposed biasing model parameters with cosmological parameters. In section \ref{model_selection} we fit the set of bias models to recent data both from observation and simulation. We present some extensions to this work in section \ref{Extensions} including the dependence of the proposed parameterisation on halo mass, and conclude in section \ref{conclusions}. In Appendix A we present a derivation of Fry’s redshift dependent galaxy bias which is both simpler and more general than in \cite{Fry1996}.

\section{Selected Bias Models}
\label{bias:models}

\begin{figure}
		\centering
		\includegraphics[scale=0.5]{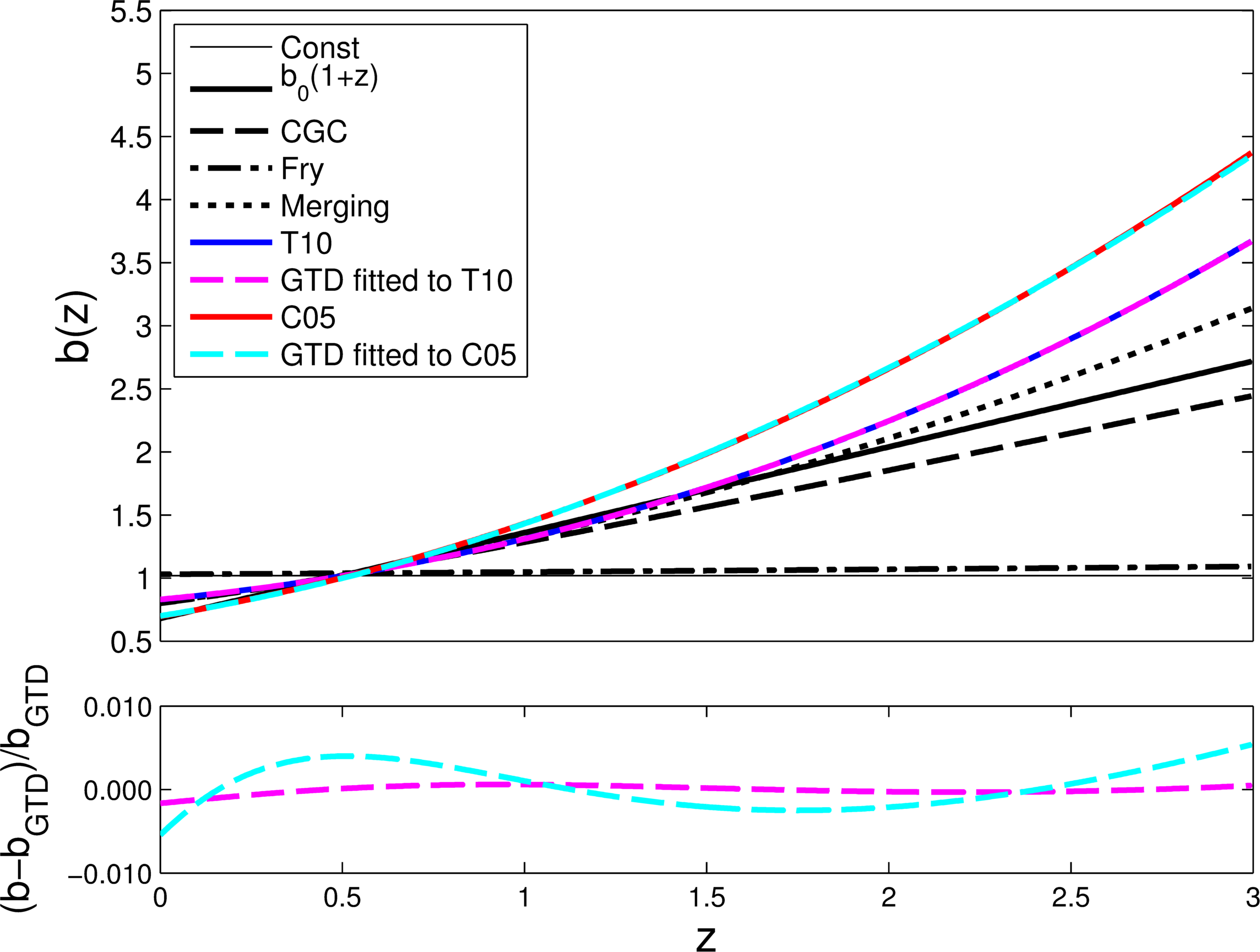}

		\caption{\label{fig:bofz_fidvals} \textit{Upper panel}: the set of redshift dependent bias models considered in this analysis, in the redshift range used in the Fisher matrix forecasts ($0<z<3$). The bias models are fitted to data from Jullo et al. (2012) (J12 hereafter) which is in the range $0.22<z<0.97$, so at higher redshifts $b(z)$ for the different models diverge. The GTD model encapsulates four of these bias models, and we show that it can very well approximate the fitting functions T10 and C05. The solid blue and solid red lines show the T10 and C05 models respectively. The dashed magenta line which sits on top of the blue T10 line is the GTD model with its three parameters tuned to T10, this fits within $<0.2\%$. The magenta line shows the GTD model fitted to the C05 model, and agrees to within $<0.6\%$. \textit{Lower panel}: fractional differences of T10 and the GTD fit to it (magenta), and of C05 and the GTD fit to it (cyan).}
\end{figure}

To investigate the impact of the choice of galaxy biasing model on parameter forecasts, we consider the models listed below. As discussed above we restrict ourselves to the redshift evolution of linear galaxy bias.Note that the evolution of bias we consider is the physical evolution with redshift seen for example in a volume limited sample, and not the apparent evolution that can be caused by selection effects. We do not exhaustively consider all parameterisations of $b(z)$ but have focused on a set of physical and empirical models, and will later contrast them with a binned function of redshift. The models listed range from the simplest possible choice, a constant bias, to more complex models deriving from theory, simulations and from observations which account for the growth and merging of collapsed structure. Each of these models are plotted as a function of redshift in figure \ref{fig:bofz_fidvals}. Note that they can all be expressed by the `Generalised Time Dependent' (GTD) parameterisation, described at the end of this section.

\begin{table*}
\scriptsize
\centering
\begin{tabular}{|c|c|c|c|c}
\hline
Bias Model & $b(z)$ & Source & Fiducial Values & Comments\\
\hline
Constant bias & $b_0$ &e.g. \cite{Peacock1994} & $b_0 = 1.02$& `Unbiased' \\
Linear Redshift Evolution & $b_0(1+z)$ &e.g. \cite{Ferraro2014}  & $b_0 = 0.68$ & \textit{Ad hoc} \\
Constant Galaxy Clustering (CGC) & $b_0/D(z)$ &  \cite{Lahav2002} & $b_0 = 0.80$ & Empirical \\
Fry & $1 + (b_0 - 1)/D(z)$   & \cite{Fry1996} & $b_0 = 1.03$ & Theoretical \\
Merging &  $0.41 + (b_{0} - 0.41)/D(z)^{\alpha}$  & \cite{Matarrese1997}& $b_0 = 0.84$, $\alpha = 1.73$ & Theoretical \\
T10 & $\nu(z)=\nu_0/D(z)$ & \cite{Tinker2010} & $\nu_0 = 0.83$ & Fitting function, N body\\
C05 & $b_0[0.53+0.289(1+z)^2]$ & \cite{Croom2005} & $b_0 = 0.85$ & Fitting function, QSO \\
Generalised Time Dependent (GTD) & $c + [b_0 - c]/D(z)^{\alpha}$ & - & $c = 0.57, b_0 = 0.79, \alpha = 2.23$ &\\
\hline
\end{tabular}
\caption{\label{table:bias_models} The set of bias models used in this analysis. Fiducial values of model parameters are fitted to J12. The resulting $b(z)$ for each model is shown in figure 1. Note that $b(z)$ for T10 is obtained by inserting the expression in the table for $\nu(z)$ into equation \ref{eq:Tinker}.}

\end{table*}

\subsection{Constant Bias Model}
\label{models:const}
The simplest possible model is that of a constant galaxy bias:

\begin{equation}\label{eq:linear_bias_deltas}
	\delta_{g}(\mathbf{x}) = b \delta_{m}(\mathbf{x}),
\end{equation}
which relates fluctuation in mass density $\delta_{m}$, and fluctuation in galaxy number density $\delta_{g}$ at location $\mathbf{x}$. This linear, deterministic relation is known to be too simplistic since galaxy bias evolves with scale and redshift at the least, but it is often favoured for ease of use.

\subsection{Linear Redshift Evolution}
\label{models:const}

A simple \textit{ad hoc} model used in the literature (e.g. \cite{Ferraro2014}) is

\begin{equation}\label{eq:linear_evolution}
	b(z) = b_0(1+z).
\end{equation}
This has no physical motivation but simply reflects the known increase of bias with redshift. (Note that this is equivalent to the CGC model below in an Einstein de Sitter Universe, however it is used here purely as a simple representation of an increasing $b(z)$).

\subsection{Constant Galaxy Clustering (CGC) Model}
\label{models:CGC}

We know from simulations (e.g. \cite{Kauffmann1999}, \cite{Somerville1999} that while the clustering of dark matter evolves, galaxy clustering is fairly constant (at $0<z<0.5$). If we assume then that the galaxy auto correlation function, $\xi_{gg} = const$, since $\xi_{mm}$ grows with $D(z)^2$ in the linear regime we can say that $\frac{\xi_{gg}(z)}{\xi_{mm}(z)} \propto \frac{1}{D(z)^2}$, or: 
\begin{equation}\label{eq:CGC}
	b(z) = \frac{b_0}{D(z)}.
\end{equation}

\subsection{Fry (1996)}
\label{models:Fry}

Analytic models and data show that biasing evolves with both scale and time, due to the highly complex process of galaxy formation and the merging of halos into larger structures. These factors make a complete description of the biasing difficult, but \cite{Fry1996} presents a simple model which gets around this by considering the biasing after the point at which galaxies form, and assuming they then evolve under gravity `passively', i.e. without merging. The redshift evolution of the biasing is then given by:

\begin{equation}\label{eq:fry}
	b(z) = 1 + \frac{b_0 - 1}{D(z)},
\end{equation}
where $b_0$ is the bias at $z=0$ and $D(z)$ is the linear growth factor, which satisfies

\begin{equation}\label{eq:growth}
	\ddot{D} + 2H(z)\dot{D} - \frac{3}{2}\Omega_{m,0}H_0^2(1+z)^3D = 0
\end{equation}
%
Here $H(z)$ is the Hubble parameter, which for a flat Universe with evolving Dark Energy is given by

\begin{equation}\label{eq:H}
	H^2(z)/H_0^2 = \Omega_M(1+z)^3 + (1-\Omega_M)e^{3\int_ 0^z d ln(1+z')[1+w(z')]},
\end{equation}
where $w(z) = w_0 + (1-w_a)$ \cite{Chevallier2001}, \cite{Linder2003}. 

In Fry's original derivation the assumption of an Einstein de Sitter Universe is made, but a more general proof is given in Appendix A.

\subsection{Merging Model}
\label{models:merging}

The assumption in Fry's model that once galaxies have formed they make it to the present time intact is clearly unrealistic, as the accepted theory of hierarchical structure formation involves the merging of sub-units into larger structures. \cite{Matarrese1997} constructed a model which accounts for this, which we refer to as the merging model: 

\begin{equation}\label{eq:Matarrese_b_eff}
	b_{\textrm{eff}}(z) = 0.41 + [b_0 - 0.41]/D(z)^{\alpha}.
	\end{equation}
This quantity $b_{\textrm{eff}}$ is essentially an average of $b(M,z)$ over the mass range considered. The value 0.41 is given by $1-1/\delta_c$ where $\delta_c$ is the linear-theory extrapolated critical overdensity for spherical collapse, which is 1.69 for $\Omega_0 = 1$ (and weakly $z$ dependent for other cosmologies, \cite{Lilje1992}.

\subsection{Tinker \textit{et al.}, 2010 (T10)}
\label{models:Tinker}

The halo bias function of \cite{Tinker2010} was calibrated on a large set of N-body simulations spanning six decades of mass from $10^{10} h^{-1} M_{\odot}$, with halos defined such that the mean overdensity is $\Delta$ times the background. It has the useful feature of being adaptable to any value of $\Delta$, and is widely used in the literature (e.g. \cite{Bohringer2014}, \cite{Osborne2014}, \cite{Kim2013}. The fitting function,

\begin{equation}\label{eq:Tinker}
	b(\nu) = 1 - A\frac{\nu^\alpha}{\nu^\alpha+\delta^\alpha_c} + B\nu^\beta + C\nu^\gamma,
	\end{equation}
is defined in terms of $\nu (M,z)= \delta_c(z) / D(z)\sigma(M)$ where $\delta_c$ is the critical density for collapse, $\sigma(M)$ is the linear matter variance of the halo, $D(z)$ is the linear growth function and $\sigma(M)$ is the mass variance smoothed on a scale $R = (3M/4\pi\bar{\rho}_{m})^{1/3}$. The constants, calibrated on the clustering of dark matter halos in N-body simulations, are given by $A = 1.0 + 0.24y e^{−(4/y)^4}$, $\alpha = 0.44y - 0.88$, $B = 0.183$, $\beta = 1.5$, $C = 0.019 + 0.107y +0.19 e^{−(4/y)^4}$ and $\gamma = 2.4$, where $y \equiv log_{10} \Delta$ depends on the chosen definition of a halo in terms of $\Delta$. In the following we use the common value $\Delta = 200$.

In order to use this alongside the previous biasing models we need to express this $b(\nu)$ as a function of redshift. To do this, we can exploit the redshift dependence of \ $\nu(M,z)$ that comes in via the growth function (ignoring the weak redshift dependence of $\delta_c$). For halos of a given mass, the redshift dependence of $\nu$ can be expressed as 
\begin{equation}\label{eq:Tinker2}
\nu(z) = \nu_0/D(z), 
\end{equation}
where $\nu_0$ is a constant. If $\nu_0$ is determined from data, $\nu(z)$ can be calculated and put into equation \ref{eq:Tinker} to give $b(z)$. We will refer to this recasting of the \cite{Tinker2010} $b(\nu)$ as $b(z)$ as T10 hereafter.

\subsection{Croom \textit{et al.}, 2005 (C05)}

A fitting function for QSOs often used in the literature (e.g. \cite{Geach2013}, \cite{Ferraro2014} is that of \cite{Croom2005} (C05 hereafter):

\begin{equation}\label{eq:Croom}
	b(z) = b_0[0.53 + 0.289(1 + z)^2].
	\end{equation}
This  was empirically derived from the 2dF QSO Redshift Survey.

\subsection{Generalised Time Dependent (GTD) Bias Model}
\label{models:GTD}

All of the biasing models described above can be encapsulated by the following expression:

\begin{equation}\label{eq:GTD}
	b(z) = c + (b_0 - c)/D(z)^{\alpha},
\end{equation}
where $D(z)$ is the linear growth function, given by equation \ref{eq:growth}. It reproduces the first four of the models described by choosing the constants to have the following values:

\begin{itemize}

\item Constant bias: $c = 0, \alpha = 0$;
\item CGC: $c = 0$, $\alpha = 1$;
\item Fry: $c = \alpha = 1$;
\item Merging: $c = 0.41$.
\end{itemize}

The fitting functions of T10 (the \cite{Tinker2010} $b(\nu)$ recast as $b(z)$, as described in section \ref{models:Tinker}) and C05 can be very well approximated by the GTD model, as shown in figure \ref{fig:bofz_fidvals}. Here the solid blue and solid red lines show the T10 and C05 models respectively, where the constants $\nu_0$ of T10 and $b_0$ of C05 are fitted to bias measurements from \cite{Jullo2012} (J12 hereafter). Using data from their lower stellar mass bin ($10^9<M*<10^{10} h^{-1}M_{\odot}$) in the range $0.22<z<97$ the best fit value of $\nu_0$ is $0.83$, and fitting the constant $b_0$ in C05 gives $b_0=0.85$ The dashed red line which sits on top of the black T10 line is the GTD model with its three parameters tuned to T10 ($c = 0.58, b_0 = 0.83, \alpha = 2.23$). This fits within $<0.2\%$, showing that the GTD model can incorporate the halo model information encoded in T10. The magenta line shows the GTD model fitted to the C05 model ($c = 0.19, b_0 = 0.70, \alpha = 1.88$), and agrees to within $0.6\%$.


The fact that three of these models (constant bias, Fry's model and the merging model) can be encapsulated in this form has been remarked on in the literature (\cite{Matarrese1997}, \cite{Moscardini1998} but that it can also replicate T10's simulation derived galaxy bias fitting function as well as the empirical CGC and C05 models, is an important extension. 

Given that we do not know the true form of galaxy bias, the fact that the GTD parameterisation covers six well motivated models is a compelling reason to consider it. In section \ref{impact_on_forecasts} we test its performance in constraining cosmological parameters, contrasting it with a non-physically motivated binned function of redshift. In section \ref{model_selection} we then see how it fares against the other bias models when fitted to observational and simulation data.

It is worth mentioning that the GTD model, and those nested within it, involve the growth function (equation \ref{eq:growth}) which is itself cosmology dependent. So the biasing, if used as an ingredient in constraining cosmological parameters is cosmology dependent. We recommend that when using the GTD parameterisation to model galaxy bias as a nuisance parameter, the cosmology chosen for the analysis at hand is used to calculate the growth rather than fixing it with a fiducial cosmology.

\section{Survey Set Up and Forecast Method}

An overview of the modelling assumptions and methodology used to produce the Fisher forecasts of the following section are given here. For greater detail please refer to Appendix \ref{app:FM}.

\subsection{Survey Set Up}

We assume a photometric optical survey like the Dark Energy Survey (DES, www.darkenergysurvey.org), observing $\sim$300 million galaxies over 5000 deg$^2$, and consider the 2-point correlations of galaxies in five tomographic redshift bins with photometric redshift errors. We compute the matter power spectra required for the Fisher forecasts in a spatially flat $\Lambda$CDM cosmology with WMAP7 (\cite{Komatsu2011}) fiducial paramters $\Omega_m = 0.272, w_0 = -1, w_a = 0,h = 0.71,\sigma_8 = 0.809,\Omega_b = 0.0449, n_s = 0.963$. 

We assume that there exists for each galaxy a photometric redshift with error $\sigma_z = \delta_{z}(1+z) = 0.05(1+z)$. We take the galaxy redshift distribution to follow a Smail-type distribution,
\begin{equation}
n(z)=z^{\alpha}\exp{\left(- \left( \frac{z}{z_{0}} \right)^{\beta}\right)},
\end{equation}
We bin the galaxies into five tomographic redshift bins with a roughly equal galaxy number density per bin.

We conduct all calculations over a redshift range $z: 0 \rightarrow 3$ and a $k-$range $k: 0.0001 \rightarrow 50$.

\subsection{Parameter Constraints}

In this work we focus on the impact of different bias models on cosmological parameter constraints, which we determine using the Fisher matrix formalism. All calculations use the Limber approximation and we do not model Redshift-space Distortions (RSD). RSD cause an excess of clustering along the line of sight (\cite{Kaiser1987}) and so an increase in the amplitude of the angular power spectrum at large scales. This means that projected angular power spectra resulting from the Limber approximation and from an exact calculation with RSD diverge at low  $\ell$. For a DES-like photometric survey, they diverge significantly at $\ell \lesssim 50$ (\cite{Thomas2010}), so to account for the fact that we do not model RSD we discard the projected angular power spectrum below this threshold $\ell_{min} = 50$. 

Since the bias models we consider are only applicable in the linear regime we exclude the projected power spectrum above a threshold angular frequency with multipole: 

\begin{equation} \label{eqn:lmax}
\ell^{(i)}_{max} = k^{max}_{lin}(z^{(i)}_{med}) f_k (\chi(z_{med}^{(i)}))
\end{equation}
where $z^{(i)}_{med}$ is the median redshift of bin $i$, $f_k (\chi(z_{med}))$ is the comoving angular diameter distance, and we follow \cite{Joachimi2010} and use $ k^{max}_{lin} \approx 0.132\ z\ h $Mpc$^{-1}$. This results in the cuts in $k$ and $\ell$ given in table \ref{table:ellcuts} by redshift bin.

The Fisher matrix is given by:

\begin{equation}
F_{\alpha\beta} = \sum^{l_{max}}_{l=l_{min}} \sum_{(i,j),(m,n)} \frac{\partial C^{ij}(l)}{\partial p_{\alpha} }{\rm Cov}^{-1} \left[  C^{ij}(l), C^{mn}(l) \right] \frac{\partial  C^{mn}(l)}{\partial p_{\beta} }
\end{equation}
and marginalised errors on cosmological parameters are calculated as $\sigma_i = \sqrt{(F^{-1})_{ii}}$ (see Appendix \ref{app:FM} for more details). 

To compare the impact on parameter constraints of the galaxy bias different models we use the Figure of Merit (FoM) defined by the Dark Energy Task Force (DETF) (\cite{Albrecht2009}, which is equal to the inverse of the area enclosed within the 1-sigma error ellipse in the $w_0, w_a$ plane. For this we take the inverse of the full Fisher matrix, extract the rows and columns associated with $w_0,w_a$ and take the determinant:

\begin{equation}\label{eq:fom}
	FoM = \frac{1}{\sqrt{\text{det}(F^{-1})_{w_0,w_a}}}.
\end{equation}

\begin{table}
\centering
\begin{tabular}{|c|c|c|c|}
\hline
$z_{bin}$ & $z_{med}$ & $\ell_{max}$ & $k^{max}_{lin} h$ Mpc$^{-1}$\\
\hline
1 & 0.38  & 55 &  0.05\\
2 & 0.61 & 128 & 0.08 \\
3 & 0.79 &  206 &  0.10\\
4 & 1.00 &  315 & 0.13\\
5 & 1.38 &  521 &  0.18\\
\hline
\end{tabular}
\caption{Cuts in angular frequency in each redshift bin.}
\label{table:ellcuts}
\end{table}

\subsection{Shift in Cosmological Parameters Caused by Assuming an Incorrect Bias Model}
\label{shift_method}

As well as affecting the magnitude of the errors on forecasted cosmological parameters, assuming an incorrect bias model will also cause a shift in the estimated values of parameters. We extend the Fisher matrix formalism to calculate the shift in cosmological parameters produced when an incorrect bias model is assumed (\cite{Cypriano2010a}, \cite{Shapiro2010a}). The shift in each parameter is given by:
\begin{equation}
\delta p_{\alpha} = F_{\alpha\beta}^{-1} \sum_{l} \Delta C_{ij}(l) \left( {\rm Cov } \left[ C_{ij}(l),C_{mn}(l) \right] \right)^{-1} \frac{\partial C_{mn}(l)}{\partial p_{\beta}},
\label{eqn:shift}
\end{equation}
where $\Delta C(l) = C(l)_{\textrm{true}} - C(l)_{\textrm{assumed}}$, and all other terms are calculated using the assumed (incorrect) model.

\section{Results: Impact of Bias Model Choice on Parameter Forecasts}
\label{impact_on_forecasts}

In this section we investigate how important the choice of galaxy bias model is for cosmological inference, and whether we should care that there is currently no consensus on how the evolution of the bias should be parameterised. We do this firstly by quantifying the impact that the choice of bias model has on parameter forecasts, focusing on Dark Energy equation of state parameters $w_0, w_a$. Secondly we look at the shift in calculated values of $w_0, w_a$ that would result from assuming an incorrect bias model.

\subsection{Impact on Dark Energy FoM}
\label{Impact_on_FoM}

Table \ref{table:FoM} shows the Dark Energy FoM for each of the bias models described in section \ref{bias:models}. We also look at the effect of adding extra degrees of freedom to a bias model, selecting the simple linearly increasing bias model $b_0(1+z)$ which we know does not capture the complexities of the true redshift evolution of the bias. This initially has one free parameter, $b_0$, which we assign fiducial value of 0.68 by fitting to the J12 data. We construct the 2 parameter version of this model by adding two amplitude terms $b_1$ at $z=0$ and $b_2$ at $z=3$ with fiducial values given by $0.68(1+z)$. As more free parameters are added the additional $b_i$ are linearly spaced in between $b_1$ and $b_2$, and as the $b_i$ vary in the Fisher calculation the resulting $b(z)$ is interpolated with a spline. In calculating marginalised errors we include a basic Planck TT prior (Dark Energy Survey TCP group, private communication) to best illustrate the order of magnitude constraints expected from early plus late Universe cosmology with this forecast DES data.

We find that the number of parameters has a significant impact on the FoM: in going from a single amplitude term to two free parameters there is a reduction in the FoM by a factor of more than 10, from 37.41 to 3.68. This is expected since introducing additional parameters leads to degeneracy between evolution of the bias and evolution of the growth. Adding further parameters has a much smaller effect on the FoM. We find that the impact on the FoM of adding free parameters $b_i$ is of the same order whichever of the bias models listed in table \ref{table:FoM} we fit the fiducial values of these $b_i$ to: on average the FoM is reduced by a factor of 10 when a free parameter per redshift bin (i.e. five) is allowed rather than a constant bias.

We find that the choice of parameterisation also has a significant impact on the FoM. Considering for example the six bias models with a single parameter: there is a difference of 85\% in the FoM resulting from the $b_0(1+z)$ model, which gives the largest FoM (37.41) out of that subset of models, and from T10 which gives the smallest (20.16).

Although the bias models with fewest parameters are clearly preferable in terms of constraining power the problem with simple, restrictive models, as we show in the following section, is that if they do not correctly describe the true bias this leads to a shift in the values of cosmological parameters inferred



\begin{table}
\centering
\begin{tabular}{|c|c|c|}
\hline

Model & No. parameters & FoM\\
\hline
\hline
\multirow{5}{*}{$b_0(1+z)$} & 1 & 37.41\\
 & 2  & 3.68\\
 & 3  & 3.54\\
 & 5  &  3.32\\
 & 7  & 3.28\\
 \hline
Constant & 1 & 30.04 \\
CGC & 1  &  20.72\\
Fry & 1 & 24.04\\
merging & 2 &  8.24\\
T10 & 1 & 20.16\\
C05 & 1 & 33.36 \\
 \hline
GTD & 3 &3.48\\
\hline
\end{tabular}
\caption{Dark energy FoM as defined by the DETF when forecasts are calculated with different bias models for a DES-like survey, considering galaxy-galaxy 2-point clustering in five tomographic redshift bins and assuming Planck priors. The variation in FoM between models is significant, as is the degradation of the FoM caused by adding free parameters to the simple linearly increasing bias model. }
\label{table:FoM}
\end{table}



\subsection{Shift in Cosmological Parameter Forecasts Caused by Choice of Bias Model}
\label{shift}

\subsubsection{Assuming an Incorrect Bias Model}
\label{shift:wrong_model}

\begin{figure}
		\includegraphics[scale=0.4]{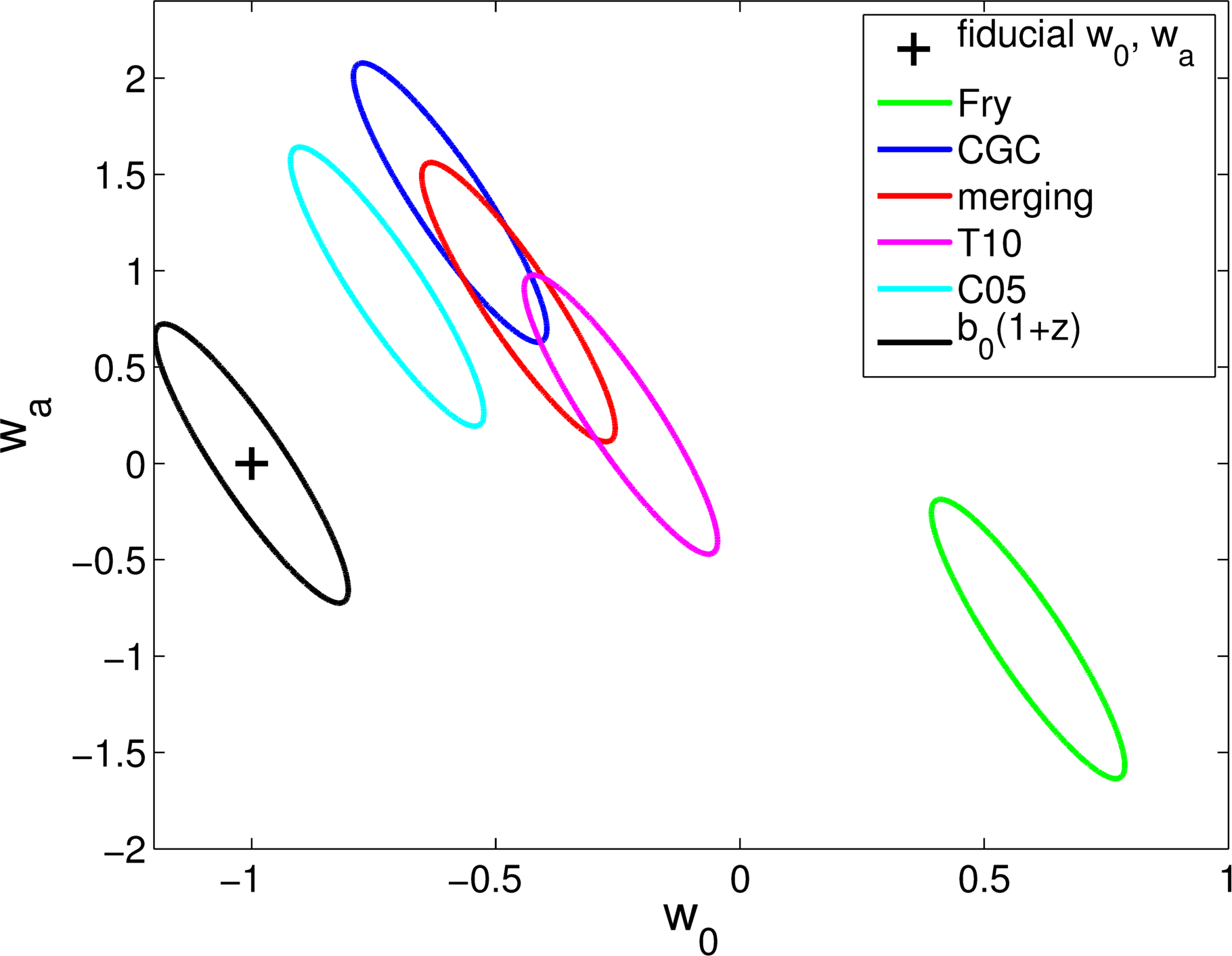}
		\caption{\label{fig:bias_demo} Shifts in measured values of $w_{0}$, $w_{a}$ caused by incorrectly modelling galaxy bias with \textit{ad hoc} model $b(z) = b_0(1+z)$, but the `true' bias is given by a different model. We consider 2-point correlations of galaxies in five tomographic bins for a DES-like survey with photometric redshift errors, and assume Planck priors. Since we do not know the form of the true bias, scenarios in which it is given by the CGC, Fry, merging, T10 and C05 models are shown. The black cross shows the fiducial values of $w_{0}$, $w_{a}$ and each ellipse shows the marginalised errors when the incorrect bias model is assumed, but the true bias is given by a different model. The offset between the centre of each ellipse and the fiducial values is the shift that would be introduced by this incorrect assumption.}
\end{figure}

Use of an incorrect bias model will cause the measured values of cosmological parameters to be shifted relative to their true values by an amount given by equation \ref{eqn:shift}. We would like to look at the size of these shifts in $w_0, w_a$ when different bias models are assumed and see whether they are significant given the statistical power of the survey. The difficulty here is that we do not know what the true bias relation is, so for the purpose of illustration we consider scenarios in which the `true' bias is described by the CGC, Fry, merging, T10 and C05 models. If we assume an incorrect bias model this will cause a shift in estimates of $w_{0}$, $w_{a}$, which is different depending on what the `true' bias looks like. Figure \ref{fig:bias_demo} illustrates the shifts caused when the true bias is given by this set of physical models, but the bias is incorrectly modelled by \textit{ad hoc} model $b(z) = b_0(1+z)$. Fiducial values of all models are fitted to J12. In figure \ref{fig:bias_demo} the black cross shows the fiducial values of $w_{0}$, $w_{a}$, and each ellipse shows the marginalised errors when this incorrect model is assumed and the true bias is given by one of the other models. The offset between the centre of each ellipse and the fiducial values of $w_{0}$, $w_{a}$ is the shift in measured values of $w_{0}$, $w_{a}$ that would be introduced by this incorrect assumption.

If additional flexibility is added to the incorrect model this shift will shrink and the marginalised errors grow due to the extra parameters. When sufficient free parameters are added the shift will become insignificant within the statistical power of the survey. We investigate how many free parameters must be added to the incorrect $b_0(1+z)$ model for this to be the case, so that an unbiased estimate of $w_{0}$, $w_{a}$ is obtained. We repeat this for a range of assumptions about what the true bias looks like (again the constant, Fry, CGC, merging, T10 and C05 bias models). We follow the methodology described in the previous section for adding linearly spaced free parameters $b_i$ to this model. When plotting marginalised statistical errors on $w_0, w_a$ we again include the Planck prior described in the previous section. This CMB prior has not been included in the calculation of the systematic shift when an incorrect bias model is assumed. This approach is taken because we want the LSS constraint itself to be unbiased without relying on degeneracy breaking from early Universe probes. This forecasting approach is broadly approximate to constraining cosmology using a DES LSS likelihood (for which we aim to produce a parameter constraint which encompasses the truth) then importance sampling our result with a CMB likelihood to reduce these statistical errors.

\begin{figure}
		\centering
		\includegraphics[scale=0.4]{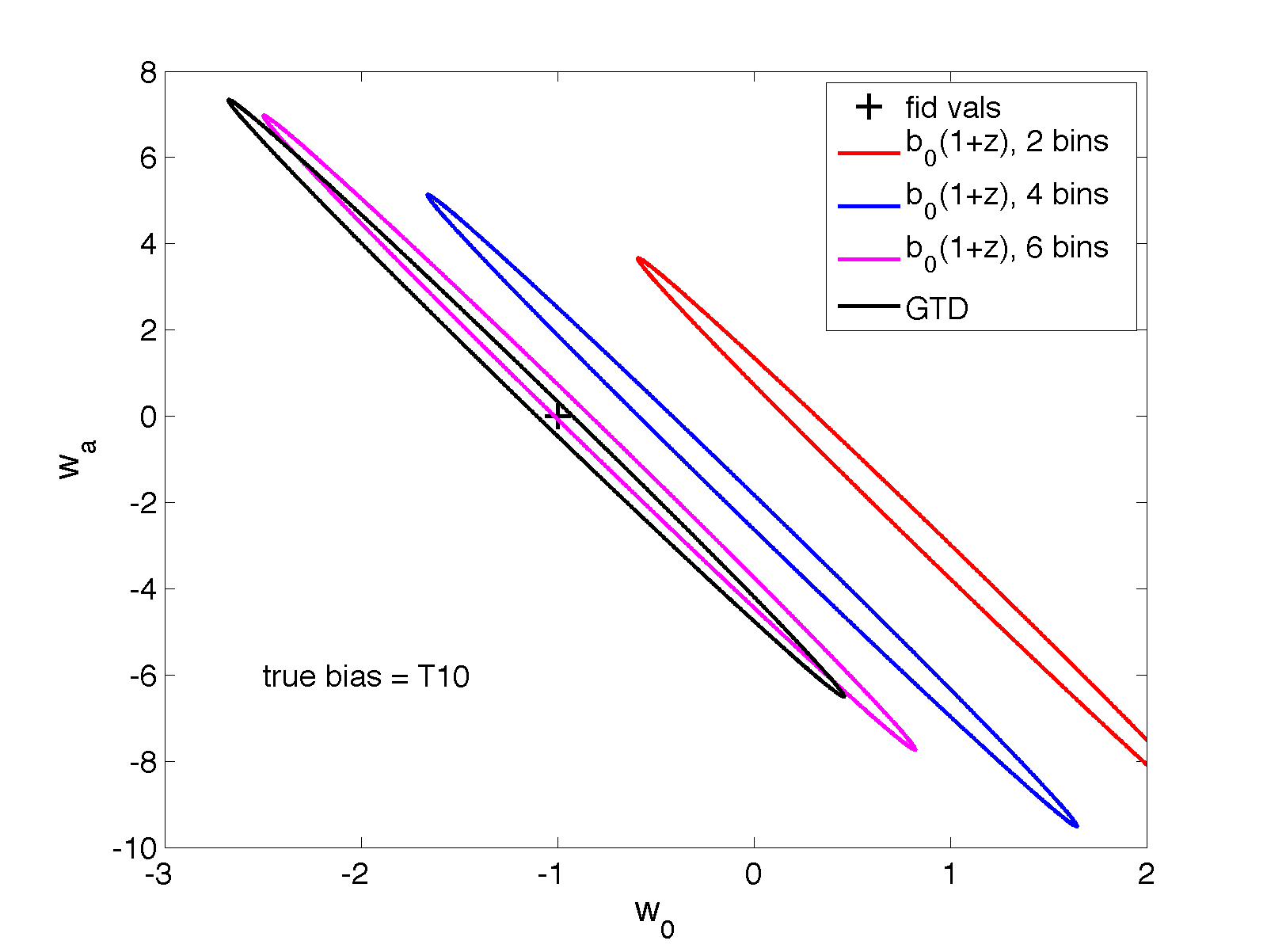}
		\caption{\label{fig:bias_assumed_1plusz_true_T10} Shift in estimates of $w_{0}$, $w_{a}$ introduced by incorrectly modelling galaxy bias with a linearly evolving function of redshift, when the `true' bias is given by the T10 model. Fiducial values of $w_{0}$, $w_{a}$ are shown by the black cross and the coloured ellipses are the 68\% marginalised errors for the assumed model, with the number of free parameters $b_{i}(z)$ increasing from two to six. The shift in parameter estimates is the offset between the cross and the centre of each ellipse, so for a model to achieve an unbiased estimate within the statistical power of the survey, the corresponding ellipse must encompass the fiducial values. It can be seen that in this scenario an unbiased estimate is achieved with six free parameters.}
		
\end{figure}

The result when the `true' bias is given by the T10 model but the linearly increasing model is used is shown in figure \ref{fig:bias_assumed_1plusz_true_T10}. The fiducial values of $w_{0}$, $w_{a}$ are shown by the black cross and the coloured ellipses are the 68\% marginalised errors for the incorrectly assumed linear $b(z)$, with the number of free parameters $b_{i}$ increasing. The resulting shift in estimates of $w_0, w_a$ is the offset between the cross and the centre of each ellipse. So for the incorrect bias model to achieve an unbiased estimate within the statistical power of the survey, the corresponding ellipse must encompass the fiducial values. In the scenario shown it can be seen that an unbiased estimate is achieved with six free parameters. 
The reason that the error ellipses are much more elongated in figure 3 than in figure 2 is that in figure 2 the assumed bias model $b0(1+z)$ has a single free amplitude; in Figure 3 more flexibility is added to model by introducing an increasing number of linearly spaced parameters. As mentioned in the previous section this leads to degeneracy between the evolution of the bias evolution of the growth.

The number of free bias parameters $b_{i}$ required to give an un-shifted estimate of $w_{0}$, $w_{a}$, for each version of the `true' bias is shown in table \ref{table:deltas}. We find that with 6 free parameters fitted to the incorrect $b(z)=b_0(1+z)$, unbiased estimates of $w_0, w_a$ can be recovered for all of the scenarios considered. For comparison we also consider the GTD model with fiducial values its three constant fitted to $0.68(1+z)$. In figure \ref{fig:bias_assumed_1plusz_true_T10} this is shown by the black ellipse, and it can be seen that it gives an unbiased estimate of $w_0, w_a$ in this case. In fact, in all of the scenarios considered the GTD model achieves an unbiased estimate with fewer nuisance parameters than the binned function of redshift, and with smaller marginalised errors (see table \ref{table:deltas}). 

\begin{table}
\centering
\begin{tabular}{|c|c|c|c|}
\hline

 Assumed Bias & `True' Bias & No. Parameters & FoM\\
\hline
\hline
\multirow{5}{*}{binned $b(z)$} &  CGC & 6 & 0.85\\
& Fry & 6 & 0.85\\
& merging & 6 & 0.85\\
& T10 & 6 & 0.85\\
& C05 & 6 & 0.85 \\

\hline
GTD & Any of the above & 3 & 0.89 \\
\hline
\end{tabular}
\caption{Number of free parameters required for an unbiased estimate of $w_0, w_a$ when one bias model is assumed, but the true bias is given by another model. Since we do not know the form of the true bias, scenarios in which the CGC, Fry, merging, T10 and C05 models describe the `true' bias are considered. In each case the bias is incorrectly modelled with a simple linear $b(z)$ with a varying number of linearly spaced free parameters, and then with the GTD model. In all cases the GTD model produces an unbiased estimate of $w_0, w_a$ with fewer nuisance parameters and a larger FoM.}
\label{table:deltas}
\end{table}

\subsubsection{Fitting Fiducial Bias to the `True' Bias}

\begin{figure}
		\centering
		\includegraphics[scale=0.4]{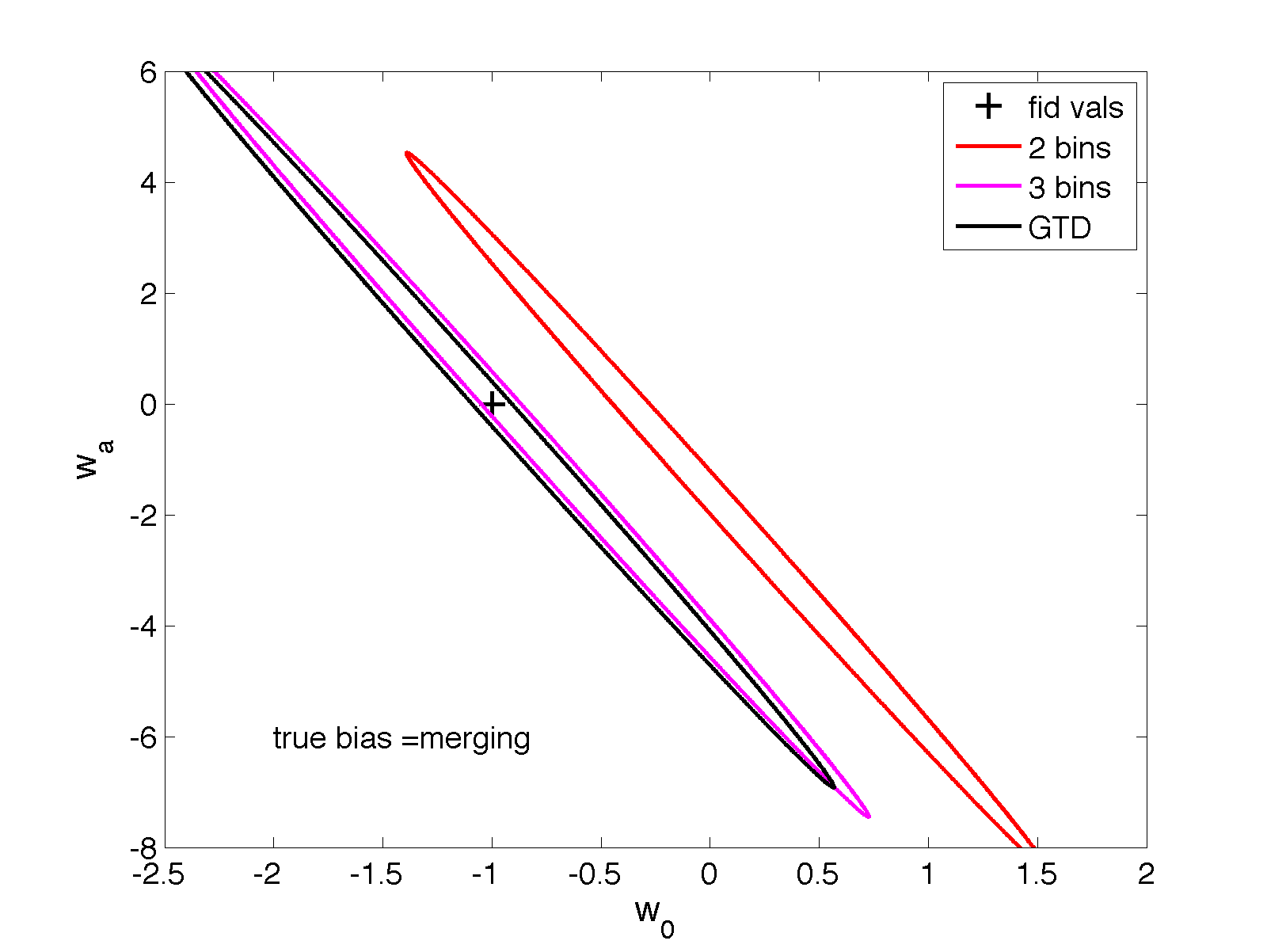}
		\caption{\label{fig:shift_grid_fittedto_true_merging} Shift in estimates of $w_{0}$, $w_{a}$ when the `true' bias is given by the merging model but it is modelled with a binned function of redshift, with fiducial values fitted to the `true' bias. Fiducial values of $w_{0}$, $w_{a}$ are shown by the black cross and the coloured ellipses are the 68\% marginalised errors for the assumed model, with the number of free parameters $b_{i}(z)$ increasing from two to three. The shift in parameter estimates is the offset between the cross and the centre of each ellipse, so for a model to achieve an unbiased estimate within the statistical power of the survey, the corresponding ellipse must encompass the fiducial values. In this scenario an unbiased estimate is achieved with three free parameters.}
		
\end{figure}

As discussed in the introduction it is common in the literature to get around the fact that we do not know which of the available bias models best describes the true bias by modelling it with a number of free parameters, which are determined by data. Here we repeat the analysis of the previous section but rather than determining fiducial values of the biasing parameters $b_i$ using a model, we fit them to the `true' bias. As in the previous section the parameters $b_i$ are linearly spaced and interpolated with a spline. The resulting shift is now caused purely by how well the binned $b(z)$ can approximate the true bias, rather than any assumptions about the form of the bias. Note that there are always five tomographic redshift bins in the calculation of the projected angular power spectrum, regardless of how $b(z)$ is binned in redshift.

Figure \ref{fig:shift_grid_fittedto_true_merging} shows the scenario in which the true bias is described by the merging model and we model it with a binned $b(z)$, with fiducial values of the binned free parameters $b_i$ fitted to the `true' bias. As expected the shifts are now smaller, and in this scenario three bins are required to recover an unbiased estimate of $w_0, w_a$. The black ellipse shows the GTD model with its three constants fitted to the `true' bias. Table \ref{table:shifts_fitted_true} shows the number of free parameters required in the binned $b(z)$ to recover an unbiased estimate of $w_0, w_a$ for the same five versions of the `true' bias as in the previous section. Three parameters are needed in all cases except when the true bias is given by Fry's model, in which case two are sufficient. For comparison we show the GTD model, with its three parameters also fitted to the `true' bias. The GTD model recovers an unbiased estimate in all cases. The two approaches to modelling the bias are comparable in terms of FoM with the GTD model leading to an improvement of the order a few percent or 23\% if the true bias is given by Fry's model.

\subsubsection{Discussion}

We have considered scenarios in which we assume a simple $b(z)$ binned in some variable number of z-bins. The fiducial values of these bins were fixed at either the `true' bias model or an incorrect bias model and the amplitude of these bins is free to vary. As more bins are used the binned function becomes more flexible but the statistical error is increased. We compare these scenarios to the GTD model which has three fixed free parameters. All scenarios show that the GTD model achieves an unbiased estimate of $w_0, w_a$ with better constraining power than a binned $b(z)$. 

One issue with the use of a binned $b(z)$, as mentioned in the introduction, is that even if it was `standardised' to the three binned free parameters that these results suggest are necessary, comparison or combination of results would be difficult as the redshift range would be specific to the survey or the analysis at hand. Additionally, a model constrained by physics will have a larger Bayesian Evidence (discussed in detail in the next section) than a more flexible model with the same number of free parameters. All of these factors support the use of the GTD parameterisation.

\begin{table}
\centering

\begin{tabular}{lllll}
\toprule 
    `True' bias & \multicolumn{2}{c}{Binned $b(z)$} & \multicolumn{2}{c}{GTD}\\
                & No. Parameters & FoM & No. Parameters & FoM   \\
\midrule
    Fry     & 2 &   2.94    &  3  &  3.61 \\
    CGC     & 3 &   3.38    &  3 &  3.46 \\
    merging & 3 &   3.44    &  3  & 3.54  \\
    T10     & 3 &   3.47    &  3  & 3.57  \\
    C05     & 3 &   3.65    &  3 & 3.75  \\            
    \bottomrule
\end{tabular}

\caption{\label{table:shifts_fitted_true} Number of free parameters required for an unbiased estimate of $w_0, w_a$ when the true bias is given by a range of models, but a binned $b(z)$ or the GTD parameterisation is assumed. No fiducial form of the bias is assumed for the binned $b(z)$, instead the fiducial values of the free parameters $b_i$ are fitted to the `true' bias. The three parameters of the GTD model are also fitted to the `true' bias in each case. The two approaches to modelling galaxy bias are comparable in terms of FoM with the GTD model leading to a FoM of the order a few percent higher, or 23\% if the true bias is given by Fry's model.}

\end{table}



\subsection{Check for Degeneracies of GTD Model Parameters with Cosmological Parameters}

\begin{figure}
	    \centering
		\includegraphics[scale=0.5]{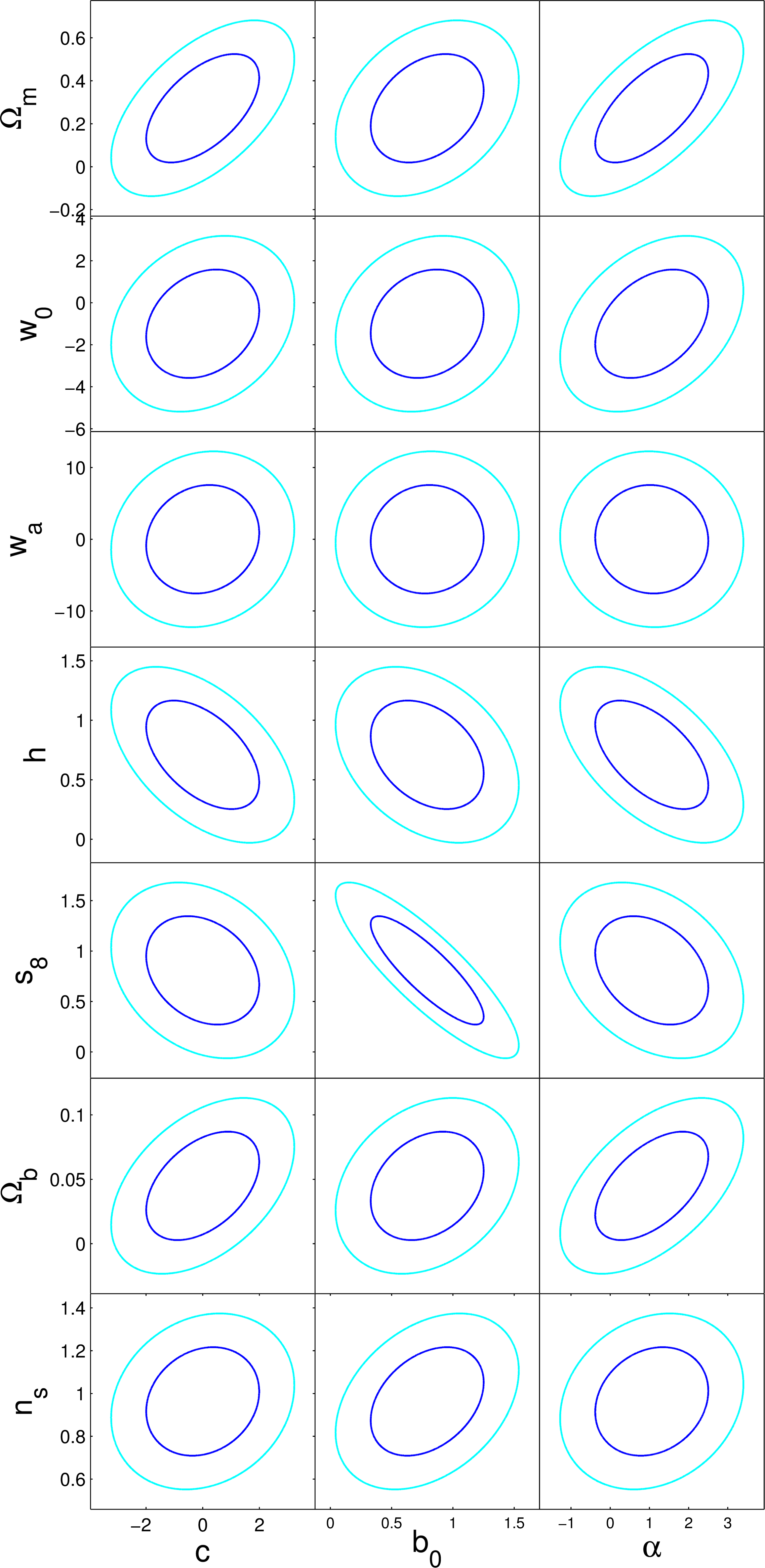}
		\caption{\label{fig:cov_gen} 68\% and 95\% marginalised errors for each parameter of the GTD bias model and the cosmological parameters used in this analysis.}
\end{figure}

Since galaxy bias is the ratio of the galaxy and matter power spectra, measurement of it is degenerate with the amplitude of the matter power spectrum, described by $\sigma_8$ (the linear-theory amplitude of the mass fluctuations). Any bias model, then, will face the issue that the amplitude of the bias at $z=0$ is degenerate with $\sigma_8$. More than one biasing parameter, allowing evolution of the bias with redshift, will mean that the bias is degenerate with evolution of the growth. In a  model with multiple parameters it is possible that they could mimic the effect of other cosmological parameters, weakening the constraints that can be placed on those parameters. To check for such behaviour figure \ref{fig:cov_gen} shows the marginalised errors of each of the parameters of the GTD biasing model and the seven cosmological parameters used in the Fisher forecasts. As expected the constant $b_0$, which controls the amplitude of the bias in the GTD model at $z=0$, is degenerate with $\sigma_8$, and less pronounced degeneracy with the matter density $\Omega_M$, the baryon density $\Omega_b$ and the expansion rate parameterised by the Hubble Parameter $h$.

\section{Model Comparison: Fitting Bias Models to Data}
\label{model_selection}

The Fisher forecasts of the previous section are a useful tool in evaluating the best approach to modelling galaxy biasing for a particular survey set up. Using this tool it has been shown in section \ref{shift} that the GTD model is a better choice to parameterise an unknown galaxy bias than a binned function of redshift. But a necessary test of its performance is to see which model fits current data best. 

\subsection{Selected Datasets}

\begin{figure}
	    \centering
	    \subfloat{
		\includegraphics[scale=0.45]{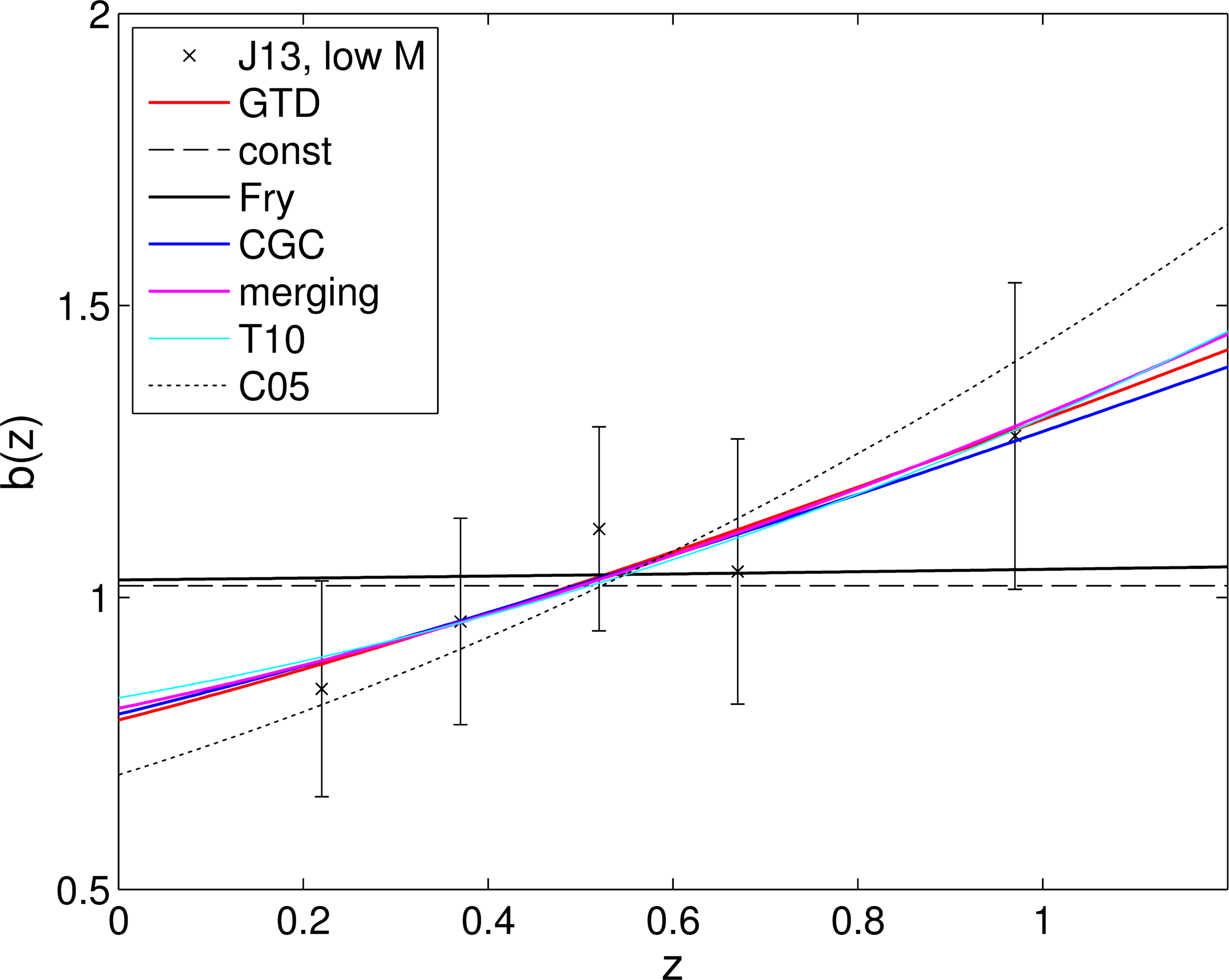}
		\label{fig:bofz_J13}}	
		
	    \subfloat{
		\includegraphics[scale=0.45]{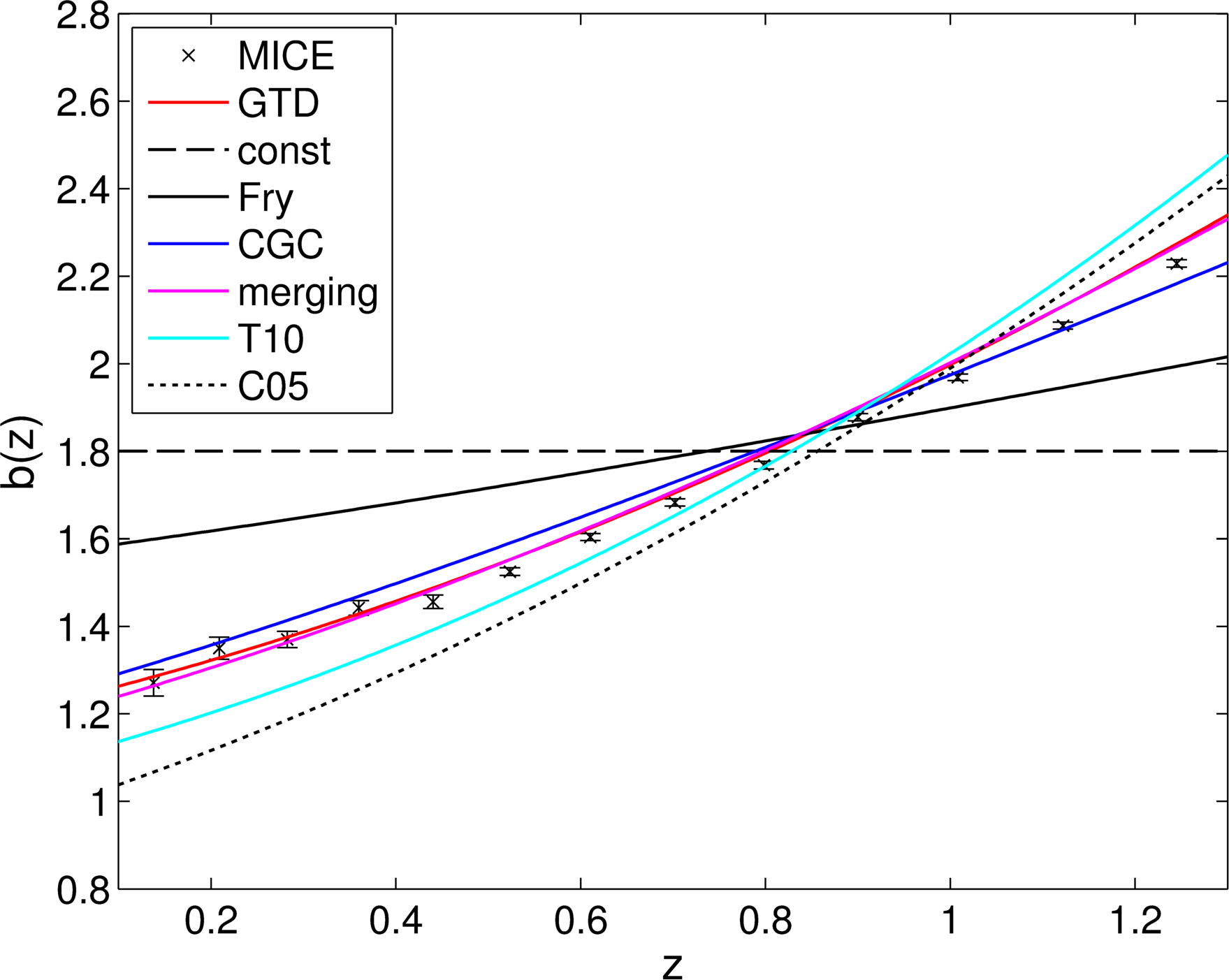}
		\label{fig:bofz_MICE_S1}}
		\caption{\label{fig:bofz_fits} Set of galaxy bias models fitted to J12's observational bias data G12's MICE simulation derived measurements of bias. In both cases the GTD model provides the best fit in terms of maximum likelihood, which is expected since the other models are either nested within it or, in the case of the T10 and C05 fitting functions, have one parameter compared the GTD model's three parameters. }
\end{figure}

In this section we fit the range of redshift dependent bias models listed in section \ref{bias:models} to measurements of galaxy bias from recent observational data and from simulation, specifically:

\begin{itemize}
\item \cite{Jullo2012} measurement of bias from weak lensing and galaxy clustering. We use data from their lower stellar mass selected samples ($10^9<M*<10^{10}$ ), where bias is computed from the ratio of galaxy to matter clustering in five redshift bins in the range $0.2 < z <1$, and averaged on scales $R>2 h^{-1}$ Mpc. 
\item \cite{Gaztanaga2012} (G12) MICE simulation derived measurements. The simulation has box size 3072 Mpc $h^{-1}$, $2048^3$ particles and redshift range $0 < z < 1.245$ (see \cite{Crocce2010} for details of the simulation). Halos are identified and fitted with a simple HOD prescription, in which the number of galaxies in a halo of mass $M_{min}$ is $1 + (\frac{M}{20M_{Min}})^\alpha$ with $M_{min} = 10^{12}M_{\odot}h^{-1}$ and $\alpha = 1$ (e.g. \cite{Scoccimarro2001}). Evolution with redshift is given by the halo mass evolution.
\end{itemize}

\subsection{Model Comparison Results}

Figure \ref{fig:bofz_fits} shows a range of bias models fitted to the J12 data (upper panel) and to the G12 results (lower panel). To compare how well the GTD model fits each of these versus the other bias models, we give the difference in maximum log-likelihoods between each bias model and the GTD model in table \ref{table:Lmax_evidence}. So a negative $\Delta$ ln$\mathop{\mathcal{L}_{max}}$ favours the GTD model. The GTD model provides the best fit in terms of maximum likelihood to both dataset. This is to be expected for models nested within the GTD model (the constant bias, Fry, CGC and merging models) and is unsurprising in the case of the T10 and C05 fitting functions, since they have a single free parameter compared the GTD model's three parameters. Several of the other models also provide a good fit to the J12 data, with the maximum likelihoods of the CGC model and the binned $b(z)$ with one or two free parameters being within 5\% of that of the GTD model. Fry's model is less flexible than the others considered and gives a poorer fit. The limiting factors of this model are that $b_0$ must be $>1$ to produce the required increase in bias with redshift, and the steepness of the evolution of $b(z)$ is very shallow unless $b_0$ becomes too large to match the bias at $z=0$. The first of these factors results in a bad fit to J12, since bias at $z=0$ of $<1$ is preferred, and the second factor causes the poor fit to G12. 


To take into account the number of parameters when comparing the models we also calculate the Bayesian Evidence, which allows one to compute the relative probability of two models given some data. For a set of parameters $\theta$, model $M$ and given data $d$, Bayes' Theorem gives the posterior probability distribution of the parameters given some data and a model:

\begin{equation} \label{eqn:bayes}
P(\theta|d,M) = \frac{P(\theta|M)P(d|\theta,M)}{P(d|M)}.
\end{equation}
The denominator is the Bayesian Evidence, $E \equiv P(d|M)$, given by 
\begin{equation} \label{eqn:bayes}
E = \int P(\theta|M)P(d|\theta,M) d\theta,
\end{equation}
which normalises the parameter posterior. Using Bayes' theorem, the ratio of the probabilities of two models $M_1$ and $M_2$, given some data is:

\begin{equation} \label{eqn:bayes}
\frac{P(M_1|d)}{P(M_2|d)} = \frac{P(M_1)P(d|M_1)}{P(M_2)P(d|M_2)}
\end{equation}
If both models are assumed \textit{a priori} to be equally likely so that their prior probabilities are $P(M_1) = P(M_2) = 1/2$, then $P(M_1|d)/P(M_2|d) = P(E_1)/P(E_2)$. The ratio of the Evidences, then, tells us the relative probability that the two models lead to the observed data. There has been discussion in the literature about the limitations of Bayesian Evidence for model comparison in Cosmology. That the effect of parameter ranges on the Evidence is significant, and that they are often difficult to define, has been discussed (\cite{Efstathiou2008}), although others are more positive regarding its usefulness (e.g. \cite{Trotta2008}).

We calculate likelihoods using a grid method with the following parameter ranges:
\begin{itemize}
\item $0<b_0<2$: based on existing data (e.g. \cite{Croom2005}, \cite{Ross2009} we do not expect the bias at $z=0$ to be higher than this, and a negative bias is not physical.
\item $0<\alpha<3$: a negative value would mean multiplying by the growth factor, which is contrary to the physical motivation of the model. A value of $>3$ results in very large bias at high redshift, although the exact choice of this cut-off is subjective.
\item $-3<c<3$: we allow negative values since there is no physical reason not to. Again the exact choice of the bounds is subjective but these choices are reasonable to prevent biases larger than expected at the redshift ranges considered ($z<1.2$). 

\end{itemize}
Flat priors are assumed within these ranges.

We quote log-Evidence ratios of each model versus the GTD model, $\Delta$ ln $E =$ln$(E_{model}/E_{GTD})$ in table \ref{table:Lmax_evidence}, so a negative $\Delta$ln$E$ indicates evidence in favour of the GTD model. If we use the common convention of the Jeffreys scale, $|\Delta$ln$E|<1$ is taken to be inconclusive; $1< |\Delta$ln$E| > 2.5$ is `substantial' evidence for the model with higher $E$; $|\Delta$ln$E|>2.5$ is `strong' evidence for that model; and $|\Delta$ln$E|>5$ `highly significant'. While these boundaries are subjective they give reasonable guidance based on the implied odds, which are $\sim3:1$ for $|\Delta$ln$E|= 1$, $\sim12:1$ for $|\Delta$ln$E| = 2.5$ and $\sim150:1$ for $|\Delta$ln$E| = 5$.

The J12 data, consisting of five data points with relatively large errors does not justify the three parameters of the GTD model. Bayesian Evidence ratios show that there is no significant Evidence to justify the additional parameters, with $1.25<|\Delta$ln$E| < 1.28$ indicating `substantial' (according to the Jeffreys scale) evidence in favour of the CGC, T10 and C05. These models provide almost as good a fit to the data as the GTD model but achieve this with one free parameter each, versus the three parameters of the GTD model.

\begin{table*}
\centering
\begin{tabular}{|c|c|c|c|c|c|c|c|}
\hline
Bias Model&DoF & \multicolumn{2}{c}{J12}  & \multicolumn{2}{c}{G12}\\
&& $\mathop{ \Delta ln\mathcal{L}_{max}}$ &$\Delta$ ln Evidence & $\Delta$ ln$ \mathop{\mathcal{L}_{max}}$ & $\Delta$ ln Evidence \\
\hline
constant    & 1 & -1.01 & 0.54 & -4085.97  & -4080.30 \\
Fry     & 1 & -0.96 & 0.34 &-1557.37   & -1551.70 \\ 
CGC         & 1 & -0.05 & 1.25  & -97.04    & -71.30  \\
Merging     & 2 & -0.06 & 0.86  & -1.73     & 0.74 \\ 
T10         & 1 & -0.08  & 1.28  & -286.88   & -278.49 \\
C05         & 1 & -0.26 & 1.10  & -522.15   & -513.97  \\
$2b_i$      & 2 & -0.03  & 0.51  & -32.26    & -28.85\\
$3b_i$      & 3 & -0.01  & -0.77 & -0.28     & -2.14\\

\hline
\end{tabular}
\caption{Model comparison results when the set of bias models in the first column are fitted to J12's measurements of bias from weak lensing and galaxy clustering and to G12's MICE simulation derived galaxy bias data. The $2b_i$ and $3b_i$ models are binned $b(z)$ with two and three free parameters. All log-likelihood ratios $\mathop{ \Delta ln\mathcal{L}_{max}}$ and log-Evidence ratios $\Delta ln$Evidence are versus the GTD model, so negative figures favour the GTD model. For J12 data: the GTD model provides the best fit, although only marginally better than the T10, CGC, merging models and the binned $b(z)$ with wither one or two free parameters, but Evidence ratios show that the data does not justify the extra parameters of the GTD model, with 'substantial' (according to the Jeffreys scale) evidence for the one parameter CGC and T10 models over GTD . For G12: again GTD provides the best fit with `strong' or `highly significant' evidence for the GTD model over all other models, except the merging model for which the evidence is 'inconclusive'.
\label{table:Lmax_evidence}}
\end{table*}

\section{Extensions}
\label{Extensions}
\subsection{Sensitivity of Results to Cut-off at Small Scales}

As discussed in the introduction this work is restricted to linear scales where the scale dependence of bias galaxy is weak (e.g. \cite{Mann1998}, \cite{Crocce2013a}, and we consider only the redshift evolution of the bias. An important extension would be to address the scale dependence of bias, which becomes important at smaller scales due to the non linear processes involved in galaxy formation. 

\begin{figure}
		\includegraphics[scale=0.45]{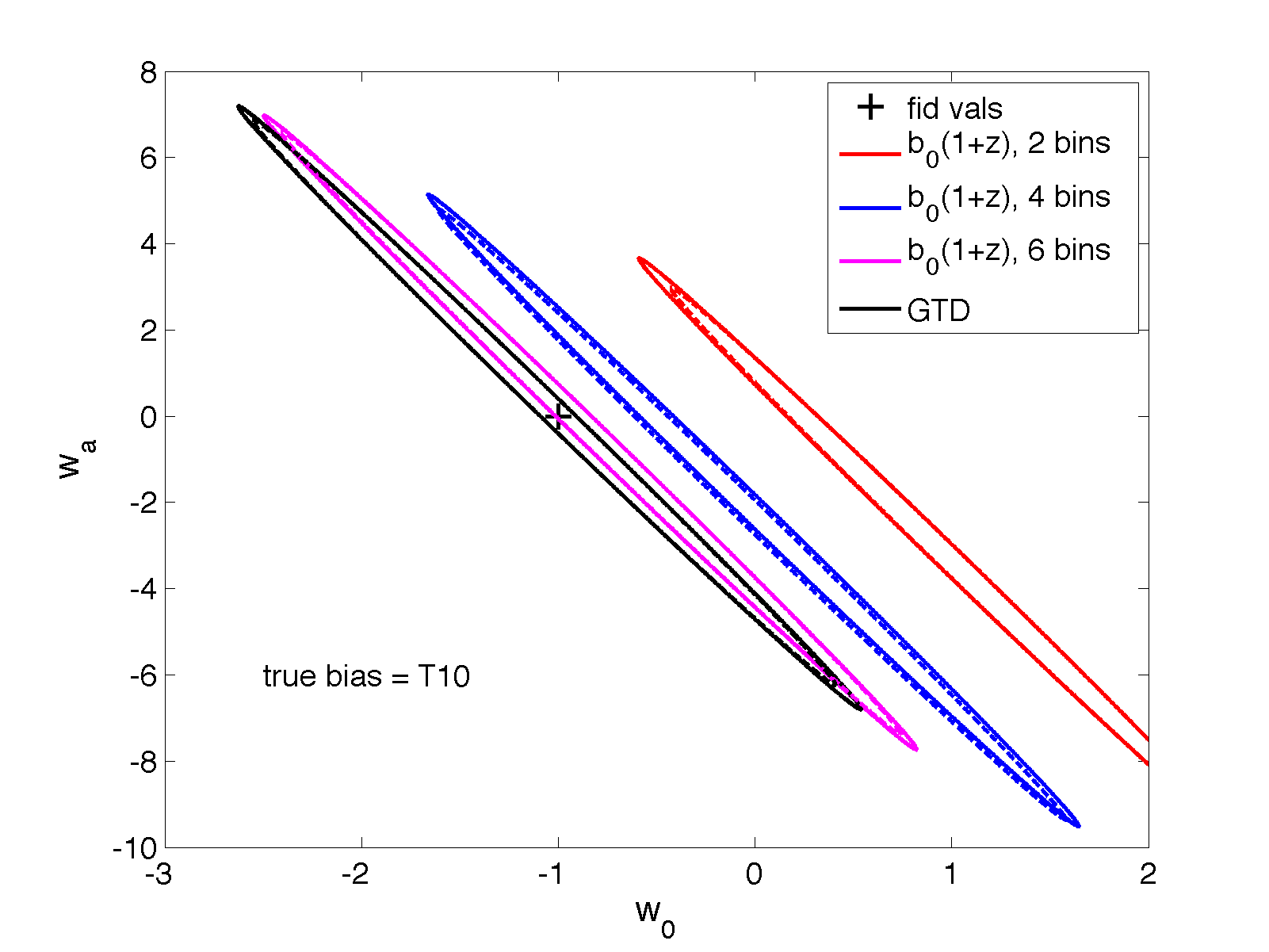}
		\caption{\label{fig:ellcut_plus10} Effect on parameter shift of a different cut to the angular power spectrum at small scales. In this scenario the `true' bias is given by the T10 model, but a linearly increasing bias with a varying number of free parameters or the GTD model is assumed. Solid ellipses show marginalised errors for the $\ell_{max}$ per redshift bin given in table 1; dashed ellipses (which lie almost on top of the solid ones) correspond to $\ell_{max}+10$. The offset between the black cross and the centre of an ellipse is the shift in estimates of $w_0, w_a$ caused by assuming a binned $b(z)$ when the true bias is given by the merging model}
\end{figure}

The cuts we make to the angular power spectrum at small scales are given by equation \ref{eqn:lmax} (with $k_{lin}^{max} = 0.18 h$ Mpc$^{-1}$ in the lowest redshift bin). To ensure that our results are not sensitive to the positioning of this cut we repeat the analysis of section \ref{impact_on_forecasts} at an angular frequency multipole $l$ of 10 higher than previously:

\begin{equation} \label{eqn:lmax_plus10}
\ell^{(i)}_{max} = k^{max}_{lin}(z^{(i)}_{med}) f_k (\chi(z_{med}^{(i)}))+10,
\end{equation}
so that mildly non linear scales are included. An example of the impact of this less severe cut of the power spectrum at small scales is shown in figure \ref{fig:ellcut_plus10}, for the scenario in which the true bias is given by the merging model. We find that the different cut-off in $\ell$ does not change the number of parameters required in a binned $b(z)$ to obtain an unbiased estimate of $w_0, w_a$, for scenarios in which the true bias is given by any of the models considered.  The important result, that the GTD model gives an unbiased estimate of these parameters with fewer nuisance parameters and smaller marginalised errors than assuming a biasing parameter per redshift bin, is unchanged. 

\subsection{Dependence of Biasing Parameters on Mass}
\label{mass_dependence}

So far we have considered only the redshift dependence of the bias, but as mentioned in the introduction, characteristics such as galaxy luminosity (see \cite{Baugh2013} for a review), colour and spectral type (e.g. \cite{Magliocchetti1999c}, \cite{Zehavi2005}, \cite{Zehavi2011}, \cite{Cresswell2009a}) also affect the biasing relation. These properties depend on the environment and also encode information about the formation and evolution of the galaxy, and it is well known that different populations trace the underlying matter density differently. More luminous galaxies, for example, tend to reside in more massive dark matter haloes and to be more strongly clustered, and so have a higher bias; red galaxies are also more biased than blue, and ellipticals more so than spirals. More massive objects form in the peaks of the galaxy density field and so are strongly clustered, and therefore more biased relative to the underlying dark matter (\cite{Kaiser1984}, \cite{Bardeen1986}).

While the GTD biasing relation does not have explicit mass dependence, since we know that the degree of biasing changes with halo mass one would expect the coefficients of the GTD model to be sensitive to this. In this case the GTD model could be fitted separately to data from different halo mass bins, and the mass dependence of bias crudely accounted for.

\begin{figure}
	    	\centering
		\includegraphics[scale=0.4]{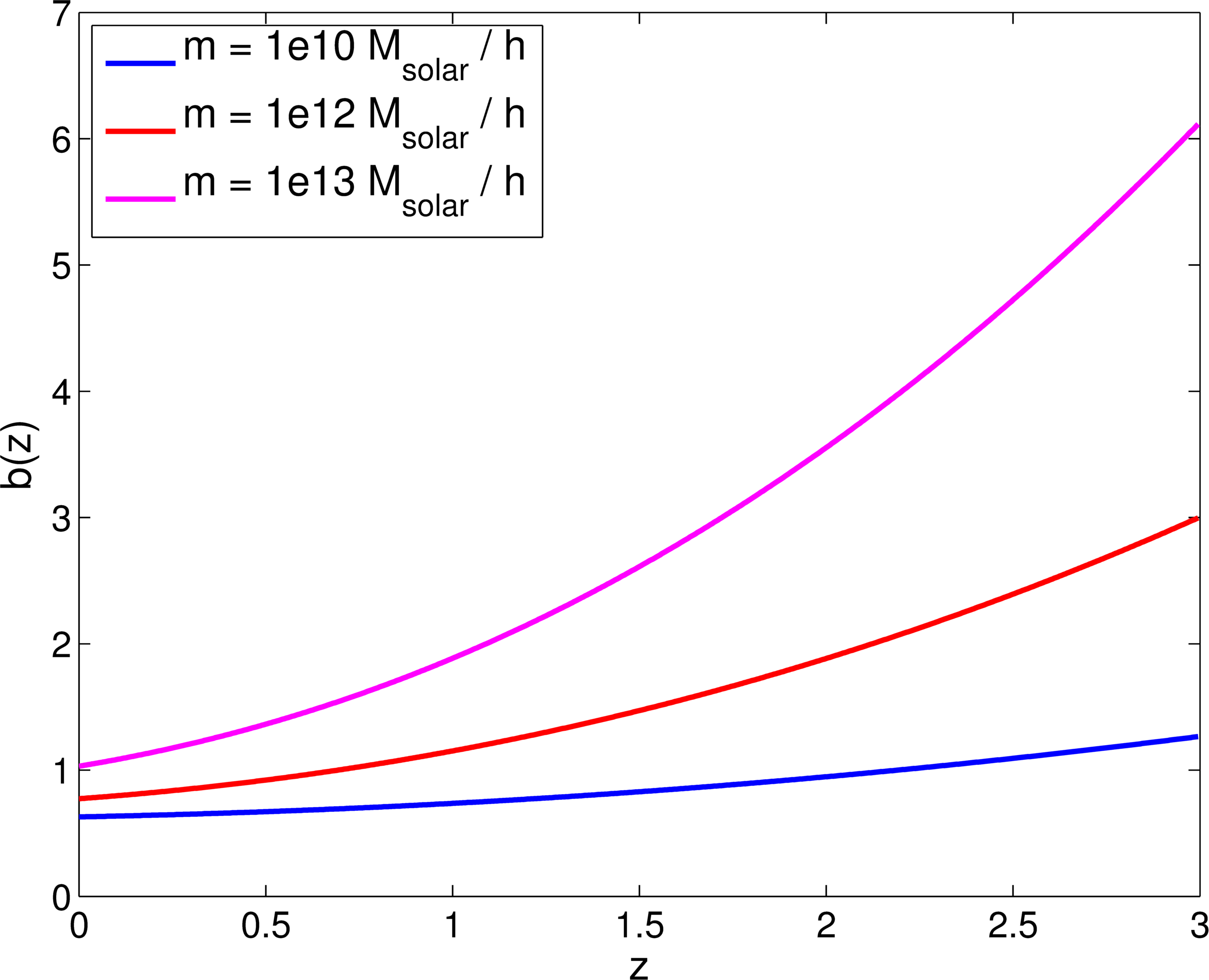}
		\caption{\label{fig:bofz_with_mass} Redshift evolution of biasing for a range of halo masses, given by the T10 bias model for $b(\nu)$ recast in terms its redshift depence as described in section 2.6. More massive objects are more biased, as expected since they form in the peaks of the density field and are more strongly clustered than lower mass objects.}
\end{figure}

To investigate this we fit the GTD model to the bias evolution, given by T10, corresponding to a range of halo masses. As described in section \ref{models:Tinker} the original \cite{Tinker2010} relation, equation \ref{eq:Tinker}, gives bias as a function of `peak height' in the linear density field, $\nu = \delta_c/\sigma_M$, where $\delta_c$ is the critical density for collapse and $\sigma_M$ is the linear matter variance at smoothing scale $R = (3M/4\pi\bar{\rho}_{m})^{1/3}$. As per section \ref{models:Tinker} this $b(\nu)$ can be recast as a function of redshift by modelling the peak height as $\nu(z) = \nu_0/D(z)$, where $\nu_0$ is the peak height $\nu$ at $z = 0$ and $D(z)$ is the linear growth function. This transforms equation \ref{eq:Tinker} into a function of redshift for a given value of $\nu_0$, or equivalently for a given halo mass at $z = 0$.  Using this relation we calculate $b(z)$ for a range of halo masses from $10^{10}h^{-1}M_{\odot}$ to $10^{13}h^{-1}M_{\odot}$ for $0<z<3$,  and fit the three parameters of the GTD model to these curves. Figure \ref{fig:bofz_with_mass} shows examples of the bias evolution for halo masses in this range. As expected this shows a greater degree of biasing for more massive objects since they form in the peaks of the density field and are more strongly clustered than lower mass objects. The GTD model fits the T10 bias evolution to within $0.3\%$ in this halo mass and redshift range. Note that the T10 $b(\nu)$ desribes halo bias, so in comparing this with the GTD galaxy bias we make the assumptions that the mass function is flat in the region considered, and that there is one galaxy per halo. The first assumption is fair since in the region considered ($M < 10^{13}h^{-1}M_{\odot}$) the mass function is fairly flat (\cite{Sheth1999}, \cite{Tinker2008}). The second assumption holds since in the linear regime the majority of the signal comes from the clustering of central galaxies.

\begin{figure}
	    	\centering
		\includegraphics[scale=0.4]{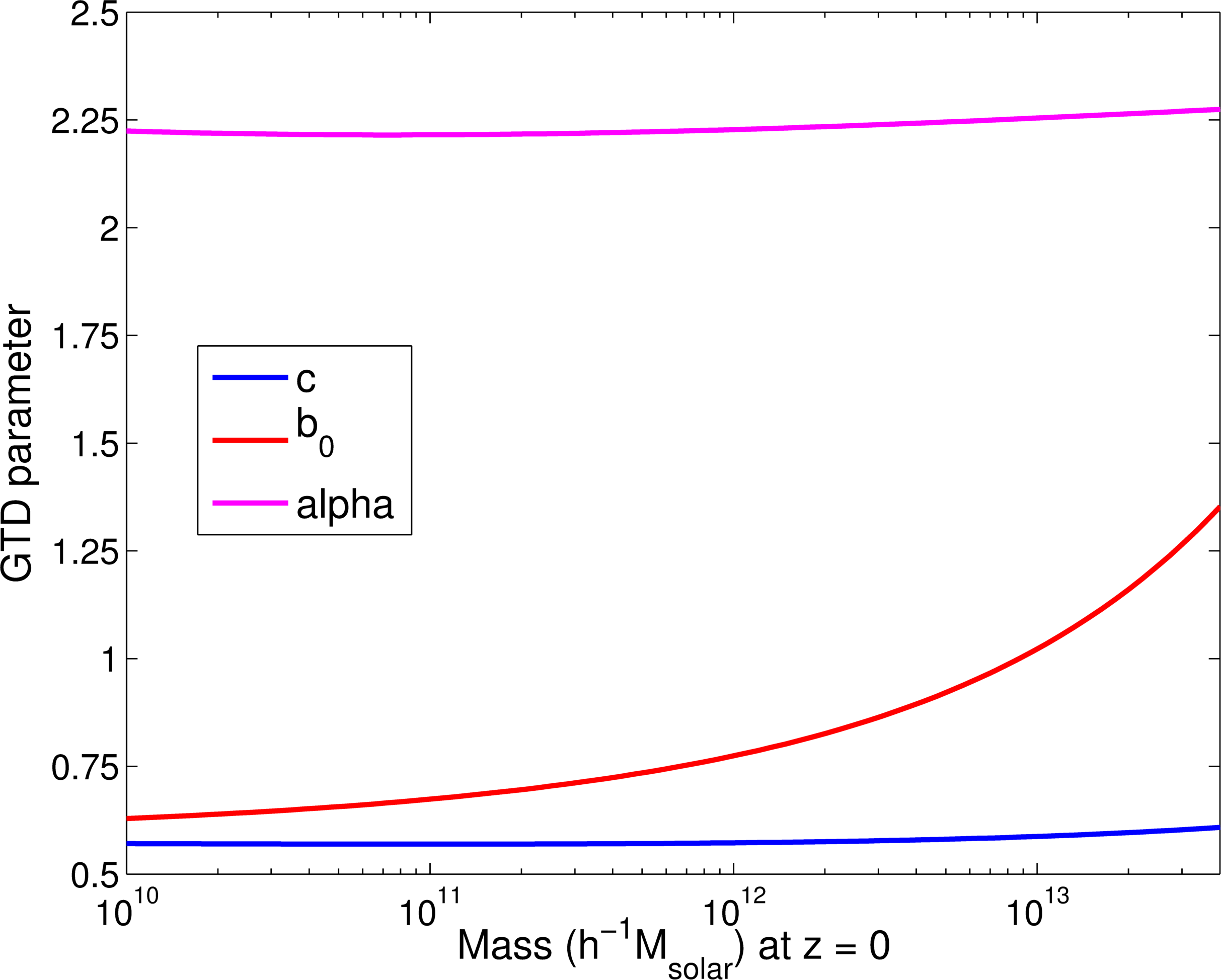}
		\caption{\label{fig:GTD_params_with_mass} Variation with halo mass of the three parameters of the GTD galaxy bias model, $\alpha, b_0$ and $c$.}
\end{figure}

 Figure \ref{fig:GTD_params_with_mass} shows the variation of the three GTD parameters $\alpha$, $b_{0}$ and $c$ with halo mass. It can be seen that $b_0$ is very sensitive to halo mass and $\alpha$ and $c$ are weakly dependent on it. Each of the GTD parameters as a function of halo mass can be described by the fitting function: $\theta_{GTD} = p + qM^n$, where $\theta_{GTD} = \alpha, b_0$ or $c$ and $M$ is the halo mass in units of $M_{\odot}$, with constants $p,q,n$ taking the values given in table \ref{table:PQ}. These provide fits to the evolution of the GTD parameters with halo mass to within $<1\%$ in mass range $M = 5\times10^{10} \rightarrow 10^{13} h^{-1}M_{\odot}$ over the redshift range $0<z<3$. This demonstrates that the dependence of bias on halo mass, although not explicit in the GTD model, is implicitly captured within these three parameters - particularly $b_0$. Galaxy samples split by halo mass can thus be modelled separately with this relation, and the mass dependence of the bias accounted for.

\begin{table}
\centering

\begin{tabular}{|c|c|c|c|}  

\toprule 
 
 $\theta_{GTD}$  &$p$ & $q$ & $n$ \\
\midrule
$\alpha$  & 0.631   &  $3.580\times10^{-7}$  & 0.464   \\
 $b_0$     &2.20     & $3.814\times10^{-5}$   & 0.293     \\
$c$         & 0.568    & $4.619\times10^{-9}$ &  0.510  \\
 \bottomrule
\end{tabular}

\caption{\label{table:PQ}Values of the constants in the fitting formula $\theta_{GTD} = p + qM^n$, where $\theta_{GTD} = \alpha, b_0$ or $c$ and $M$ is the halo mass in units of $M_{\odot}$. These provide fits to the evolution of the GTD parameters with halo mass to within $<1\%$ in mass range $M = 5\times10^{10} \rightarrow 10^{13} h^{-1}M_{\odot}$ over the redshift range $0<z<3$.}

\end{table}


An interesting extension to this work would be to repeat the analysis of section \ref{impact_on_forecasts} with different populations of galaxies defined by mass or luminosity as well characteristics such as colour and type.

.

\section{Conclusions}
\label{conclusions}

To address the issue that the literature does not currently share one approach to modelling galaxy bias evolution in cosmological inference, we have presented a Generalised Time Dependent (GTD) parameterisation $b(z) = c + (b_0 - c)/D(z)^\alpha$. This model also addresses the gap in the literature between detailed theoretical modelling of biasing and it's often simplistic implementation in cosmological inference from data. As well as generalising four commonly used redshift dependent bias models, that of a constant bias, the Constant Galaxy Clustering model (\cite{Lahav2002}), Fry's passive evolution model (\cite{Fry1996} and the merging model (\cite{Matarrese1997}) we have demonstrated its wider applicability by showing that it can also replicate the \cite{Tinker2010}) simulation derived galaxy bias fitting function and the fitting function of \cite{Croom2005} for the evolution of Type 1 quasars. We have shown that the parameters of this model are sensitive to mass, and we plan to address the scale dependence of bias and in future work.

We have performed several tests of the GTD biasing model versus other commonly used models, with the following main conclusions.
 
\noindent\textit{Impact on FoM.}  We have shown that both the number of biasing nuisance parameters and the chosen parameterisation of bias have a significant impact on the Dark Energy FoM by considering a set of commonly used bias models. A DES-like survey with five tomographic redshift bins was assumed and galaxy-galaxy 2-point correlations considered. We found that modelling galaxy bias with a free biasing parameter per redshift bin (i.e five) gives a Figure of Merit (FoM) for Dark Energy equation of state parameters $w_0, w_a$ smaller by a factor of 10 than if a constant bias is assumed.

\noindent\textit{Shift in Inferred Parameters.} Assuming an incorrect bias model also causes a shift in the values of cosmological parameters that are inferred. For a range of reasonable guesses at what the true bias might look like, we have shown that this shift can be mitigated, within the statistical power of the survey, either by adding enough free parameters to a simple linearly evolving $b(z)$ or by using the GTD parameterisation. For the case of $w_0, w_a$, six parameters are required in the linearly evolving function of redshift to produce unbiased estimates, depending on the form of the `true' bias, whereas the GTD model achieves this with three free parameters. If instead the bias is modelled with a binned $b(z)$ with fiducial values fitted to the `true' bias, we find that three free parameters are required for unbiased estimates of $w_0, w_a$. The GTD model achieves this with the same number of parameters and slightly better constraining power, with FoM between 2\% and 23\% higher than the binned $b(z)$. One issue with the use of a binned $b(z)$, as mentioned in the introduction, is that even if it was `standardised' to the three binned free parameters that these results suggest are necessary, comparison or combination of results would be difficult as the redshift range would be specific to the survey or the analysis at hand.


\noindent\textit{Degeneracies.} We have shown the marginalised errors for the parameters of the GTD model, $c, b_0$ and $\alpha$ and the seven cosmological parameters used, $\Omega_m,w_0,w_a,h,\sigma_8,\Omega_b,n_s$. As one would expect the constant $b_0$, which gives the amplitude of the bias at $z=0$, is degenerate with the amplitude of the matter fluctuations, $\sigma_8$ and less pronounced degeneracies of $c$ and $\alpha$ with matter density $\Omega_M$ and the expansion rate parameterised by Hubble parameter $h$.
 
\noindent\textit{Model Selection.} We fitted the set of bias models considered in this work to recent galaxy bias data and simulation from the literature: the \cite{Jullo2012} (J12) measurements of bias from weak lensing and galaxy clustering for lower stellar mass bin; and the \cite{Gaztanaga2012} (G12) MICE simulation derived measurements. The GTD model provides the best fit in terms of maximum likelihood to both datasets. The J12 data, consisting of 5 data points with large errors does not justify the three parameters of the GTD model, with Bayesian Evidence ratios showing that there is no significant evidence to justify the additional parameters. The G12 data shows `strong' or `very significant' evidence (according the Jeffreys scale) for the GTD model over all other models with the exception of the merging model, for which there is inconclusive evidence.


\noindent\textit{Dependence on Mass.} We have shown that the parameters of the GTD biasing model are dependent on halo mass and provided a simple fitting function for GTD model parameters $\alpha, b_0$ and $c$ in terms of halo mass. This demonstrates that the dependence of bias on halo mass, although not explicit in the GTD model, is implicitly captured within these three parameters - particularly $b_0$. Galaxy samples split by mass can thus be modelled separately with this relation, and the mass dependence of the bias accounted for.

We believe that the results in this paper recommend the use of the GTD model of galaxy bias. It has a number of positive features:
\begin{itemize}
\item The GTD model is based on physics, incorporating the best motivated z-dependent behaviour in the literature.
\item The GTD bias is easy to compute and independent of survey design.
\item Three free parameters makes parameter estimation computationally tractable while, as we have shown, producing unbiased cosmological estimates for the precision of current and future surveys. 
\end{itemize}

In some applications a particular, survey-dependent parameterisation will be preferred. This has the problem that different results in literature become difficult to compare. In these circumstances we believe that the parallel implementation of the GTD model is a simple way to produce results that can be compared consistently for very little extra effort. In future work we intend to treat dependence on scale, luminosity and galaxy type.

\section*{Acknowledgements}
We would like to thank \'{E}ric Jullo for providing the galaxy bias data from \cite{Jullo2012}, and Andrew Pontzen for helpful comments.

\appendix 
\section{Alternative Derivation of the Fry (1996) Model of Galaxy Bias} \label{Fry:proof}

In \cite{Fry1996} this biasing model is derived starting from the continuity equation, and with the limiting assumption of an Einstein de Sitter Universe. A simpler derivation, which makes no assumption about the cosmology, is presented here. 

The model assumes that once galaxies have formed they survive to the present time, with no merging of subhalos into larger structures. From this it follows that the matter and galaxy densities evolve together:

\begin{equation}\label{}
	\dot{\delta_g} = \dot{\delta_m} = const,
\end{equation}
\begin{equation}\label{}
	\delta_{g}(0) - \delta_{g}(t) = \delta_{m}(0) - \delta_{m}(t).
\end{equation}
Dividing throughout by $\delta_{m}(0)$:
\begin{equation}\label{}
	\frac{\delta_{g}(0)}{\delta_{m}(0)}-\frac{\delta_{g}(t)}{\delta_{m}(0)} = 1 - \frac{\delta_{m}(t)}{\delta_{m}(0)}
\end{equation}
where the first term is $b_0$, the second can be written as $b(t) D(t)$, where $D(t)$ is the linear growth function, and the final term is $D(t)$. We then have
\begin{equation}\label{}
	b(t) = 1 + \frac{b_0 - 1}{D(t)},
\end{equation}

which is the model presented in \cite{Fry1996}, but now applicable for any cosmology.

\section{Fisher Matrix Forecasts}
\label{app:FM}

 We forecast results using the projected angular power spectra ($C_l$s) framework (e.g. \cite{Peebles1973}, 1994, \cite{Blake2007a}, \cite{Thomas2010}. The galaxy auto-correlation, $C^{ij}(l)$ is calculated between tomographic bins, and the Limber approximation is assumed (e.g. \cite{Joachimi2010}):
 
 \begin{equation}
	C^{ij}(l) = \int \frac{d\chi}{\chi^2} n^{i}(\chi) n^{j}(\chi) b_{g}^2 P_{\delta}(k,\chi).
 \end{equation}
Here $b_g$ is galaxy bias, $\chi$ is the comoving distance, $i,j$ denote tomographic bins, $n(\chi)$ is the galaxy redshift distribution, $P_{\delta}(k,\chi)$ is the nonlinear matter power spectrum, calculated from the Eisenstein \& Hu fitting formula and the Halofit NL fitting function.
We produce forecasts using the Fisher Matrix formalism, 
\begin{equation}
F_{\alpha\beta} = \sum^{l_{max}}_{l=l_{min}} \sum_{(i,j),(m,n)} \frac{\partial C^{ij}(l)}{\partial p_{\alpha} }{\rm Cov}^{-1} \left[  C^{ij}(l), C^{mn}(l) \right] \frac{\partial  C^{mn}(l)}{\partial p_{\beta} }
\end{equation}
where $\alpha,\beta$ denote pairs of varied cosmological (or nuisance) parameters. Our default parameter set is
\begin{equation}
p_{\alpha} = \{ \Omega_m,w_0,w_a,h,\sigma_8,\Omega_b,n_s,\vec{b}\},
\end{equation}
where $\vec{b}$ denotes the nuisance parameters appropriate to the model of galaxy bias under consideration. 

The covariance matrix is given by
\begin{equation}
Cov[C_{\ell}^{ij},C_{\ell}^{mn}] =\frac{ \hat{C}_{\ell}^{im}\hat{C}_{\ell}^{jn}+\hat{C}_{\ell}^{in}\hat{C}_{\ell}^{mj}}{f(\ell)f_{sky}} 
\end{equation}
with
\begin{equation}
\hat{C}_{\ell}^{ij} \equiv C_{\ell}^{ij} + \delta_{ij}1/\bar{n_i}  
\end{equation}
in which the second term accounts for shot noise, where $\bar{n_i}$ is the average number density of galaxies per steradian in bin $i$ and 

\begin{equation}
f(\ell) \equiv \sum_{\ell'=\ell_{min}(\ell)}^{\ell_{max}(\ell)}(2\ell'+1).
\end{equation}

Marginalised errors on individual cosmological parameters can then be calculated as $\sigma_i = \sqrt{(F^{-1})_{ii}}$.


\bibliographystyle{mn2e}
\bibliography{bias_paper}

\end{document}